\documentclass[aps,prfluids,preprint,superscriptaddress,floatfix,showkeys]{revtex4-2}

\usepackage[normalem]{ulem}
\usepackage{graphicx}
\usepackage{xcolor}
\usepackage{hyperref}
\hypersetup{
    colorlinks = true,
    urlcolor   = blue,
    citecolor  = blue,
    linkcolor  = blue,
}
\usepackage{siunitx}
\usepackage{amsmath}
\usepackage{amssymb}
\usepackage{bm}
\usepackage{stmaryrd}  
\usepackage{enumitem}
\usepackage{algorithm}
\usepackage{algpseudocode}
\usepackage{comment}

\begin{document}

\title{Electrokinetic Effects on Flow and Ion Transport in Charge-Patterned Corrugated Nanochannels}

\author{Thomas Petersen}
\email{thomasp3@usc.edu}
\affiliation{Sonny Astani Department of Civil and Environmental Engineering, University of Southern California, Los Angeles, California}
\affiliation{Department of Aerospace and Mechanical Engineering, University of Southern California, Los Angeles, California}

\author{Pouya Golchin}
\affiliation{Sonny Astani Department of Civil and Environmental Engineering, University of Southern California, Los Angeles, California}

\author{Jinwoo Im}
\affiliation{Earth \& Environmental Sciences Area, Lawrence Berkeley National Laboratory, Berkeley, California}

\author{Felipe P. J. de Barros}
\affiliation{Sonny Astani Department of Civil and Environmental Engineering, University of Southern California, Los Angeles, California}

\date{\today}

\begin{abstract}
The interplay between mechanical and electrochemical forces is important to the bulk flow and transport of ions through varying natural, synthetic, and biological pores. In the current article, we demonstrate that the phase offset between the sinusoidal surface charge modulations and geometric undulations in a corrugated nanochannel provides a tunable mechanism for flow gating and rectified, diode-like ion transport under pressure-driven or electric-field driven conditions. To demonstrate these physics, the fully coupled and non-linear Poisson–Nernst–Planck–Stokes equations are solved numerically and reveal an underlying two-regime structure. At low driving force (Regime I), throughput is suppressed below the Poiseuille limit by a localized streaming potential that pins counterions within the electric double layer; at high driving force (Regime II), above a threshold pressure, the mechanical force overcomes electrostatic resistance, producing an abrupt, orders-of-magnitude rise in mean velocity. Electroosmotically driven flow undergoes a qualitatively similar but smoother transition. For mechanically driven flow, reversing the applied pressure gradient activates selective transport of the opposite ionic species and generates a net ionic current whose sign and magnitude are set by the charge-geometry alignment. Further investigating the ionic transport behavior shows that peak charge selectivity is achieved at near-complete electric double layer overlap and at a pressure gradient just below the Regime I--Regime II threshold. Random walk particle tracking confirms selective rectification and quantifies the dependence of ion dispersion on surface charge placement across both regimes.
\end{abstract}


\maketitle

\section{Introduction}
\label{ref:sec:introduction}

Technological advances in the control of species transport through porous materials have significantly impacted diverse applications of societal relevance. These range from solute migration in natural and engineered porous media~\cite{dentz2023review}, probabilistic risk assessment of groundwater contamination and remediation~\cite{henri2015}, and drug delivery~\cite{shipley2010}, to microfluidics~\cite{squires2005microfluidics}, heat exchange~\cite{webb1971heat}, and membrane filtration systems~\cite{sanaei2017}. At the pore scale, narrow conduits and complex structural morphologies play a decisive role in governing solute transport dynamics. Accordingly, improved understanding of transport in confined geometries is crucial for linking microscopic processes to macroscopic behavior.

Pore morphology and fluid-solid interface texture have long been exploited to modulate flow and enhance scalar transport~\cite{ling2018,Yoon_Dentz_Kang_2021,ling2024}. Periodically structured or wavy channels can augment heat and mass transfer~\cite{bolster2009PoF,marbach2019active,nishimura1995mass,patera1986exploiting}, though the relationship between geometry and solute dispersion is non-monotonic: Surface undulations may hinder rather than enhance transport in some regimes~\cite{bolster2009PoF,mohammadi2013pressure,marbach2018transport}. In nano- and micro-confined systems, flow separation, entropic barriers~\cite{reguera2001kinetic}, and surface-driven forces combine to produce counterintuitive behavior. Electrokinetic phenomena offer complementary flow control, with applications in enhanced oil recovery~\cite{thomas2008enhanced}, electro-remediation of contaminated soils~\cite{probstein1993removal,acar1993principles}, and desalination~\cite{epsztein2020towards,deng2015water,sapp2024deionization}. Studies of electroosmotic flow through charged slits~\cite{burgreen1964electrokinetic} and Coulombic interactions in porous media~\cite{rolle2018nernst,sprocati2022} have established the sensitivity of transport to electrostatic boundary conditions under advection--diffusion coupling. These same interfacial controls are central to biological pores. For example, a potassium-selective ion channel combines negatively charged entrances, a predominantly hydrophobic inner pore, and a polar carbonyl-lined selectivity filter to concentrate cations and discriminate $\mathrm{K}^+$ from $\mathrm{Na}^+$~\cite{doyle1998structure}.

Patterning surface charge in tandem with geometric features enables precise control over local flow structures. Experimentally, \citet{vlassiouk2007nanofluidic} used targeted chemical modification of a conical poly(ethylene terephthalate) nanopore to create adjoining positively and negatively charged surface regions, demonstrating that axial surface-charge patterns can be fabricated and used to produce strong ionic-current rectification. Nonuniform charge distributions induce electroosmotic eddies and recirculating flows~\cite{anderson1985electroosmosis,stroock2000patterning}, and similar patterns arise in field-driven flow encountering geometric perturbations~\cite{park2006eddies}. Combining nonuniform charge with geometric asymmetry generates directional flow under unbiased agitation, as predicted by Ajdari's linear analysis~\cite{ajdari1995electro,ajdari2000pumping} and confirmed experimentally~\cite{siwy2002fabrication}. Charge oscillations along a wall also affect the far-field bulk electrolyte: A field applied parallel to the wall generates nonuniform concentration polarization and axial gradients in the electric field~\cite{khair2008fundamental,goyal2024generalizing}.

At the nanoscale, the Debye screening length extends appreciably into the channel width or overlaps with the double layer of the opposing wall. \citet{curk2024discontinuous} identified transitions between electrostatically dominated and mechanically driven flow regimes in periodically charged flat channels, revealing gating behaviors and threshold responses. \citet{malgaretti2019driving} derived analytical expressions for corrugated channels in the linear response regime, showing that solute and solvent fluxes can be regulated by a combination of pressure, electric potential, and osmotic driving, and that ionic current increases with corrugation amplitude under osmotic forcing. \citet{shrestha2025self,shrestha2025universal} studied traveling-wave surface charge oscillations in cylindrical and slit pores, demonstrating that nonlinear coupling between surface potential gradients and the electric double layer (EDL) generates boundary-driven electrokinetic flows consistent with microelectrode array experiments~\cite{ramos1998ac,ramos1999ac,cahill2004electro}. The advent of ion-beam milling and related nanofabrication techniques~\cite{li2001ion,lanyon2007recessed} demands analysis of electrokinetic flow with large Debye lengths that preserves the full nonlinear coupling between electrostatic potential and charge distribution~\cite{bocquet2010nanofluidics}.

The PNPS equations and the numerical techniques employed here---a boundary-fitted coordinate mapping and a staggered MAC discretization---are well established; the contribution of the present work lies in the physical behavior revealed when they are applied without linearization to corrugated nanochannels with patterned surface charge. Earlier analyses of charge-patterned electrokinetic transport have employed the Debye--Hückel approximation~\cite{ajdari1995electro,ajdari1996generation}, lubrication and linear-response approximations~\cite{malgaretti2019driving}, or perturbative expansions about equilibrium~\cite{shrestha2025self}, while fully nonlinear particle-based simulations of the related gating transition have focused on flat channels~\cite{curk2024discontinuous}. Here, the mean channel width is \(W=5.25\) nm, comparable to the Debye length over much of the parameter space, so overlap of the EDLs is a central mechanism governing the flow and transport behavior. We resolve this overlap self-consistently by solving the Poisson equation together with the Nernst--Planck ion-density fields and the Stokes momentum balance in the physical corrugated geometry.

Building on the frameworks of \citet{ajdari2000pumping} and \citet{curk2024discontinuous}, we examine three configurations: electric field-driven flow with spatially alternating surface-charge polarity, pressure-gradient-driven flow with spatially alternating surface-charge polarity, and, for comparison, electric field-driven flow with single-polarity surface charge. In the first configuration, the electric field produces no net axial force in a flat channel, whereas corrugation breaks the symmetry of the local recirculation zones and generates axial flow. The pressure-driven configuration isolates the effects of charge--geometry alignment on species transport, while the single-polarity configuration places the Regime I--Regime II transition in the context of the gating behavior identified by \citet{curk2024discontinuous} for flat channels with surface charge patches of a single polarity.

The principal physical contributions are fourfold. First, we show that the phase offset between the surface-charge pattern and the wall corrugation controls the sign and magnitude of the ionic current, producing rectified, diode-like transport under purely pressure-driven conditions. Second, we identify two flow regimes separated by a sharp transition governed by the competition between the applied driving force and the electrostatic force acting on the charge patches. Third, comparison with linear-response theory demonstrates that the Debye--Hückel approximation can underestimate charge densities and velocities by up to an order of magnitude in the strongly overlapping-EDL regime where these effects arise. Finally, random walk particle tracking~\cite{rizzo2019par2} reveals how the two ionic species separate, become trapped by or released from individual charge patches, and approach species-dependent long-time Fickian transport. Together, these results demonstrate that bulk flow and ionic transport can be substantially decoupled. In the pressure-driven configurations, the mean flow rate is governed primarily by the applied pressure gradient, whereas ionic fluxes and dispersion remain highly sensitive to the charge--geometry phase offset.

The remainder of the paper is organized as follows. Section~\ref{sec:model} introduces the governing PNPS equations, the nondimensional parameter space, and the numerical approach. Section~\ref{sec:analysis_discussion} presents the results, beginning with validation against linear-response theory and followed by analyses of the flow regimes, ionic fluxes, and particle-transport statistics. Section~\ref{sec:conclusion} summarizes the conclusions and discusses potential extensions.

\begin{figure}[t]
    \centering
    \includegraphics[width=\linewidth]{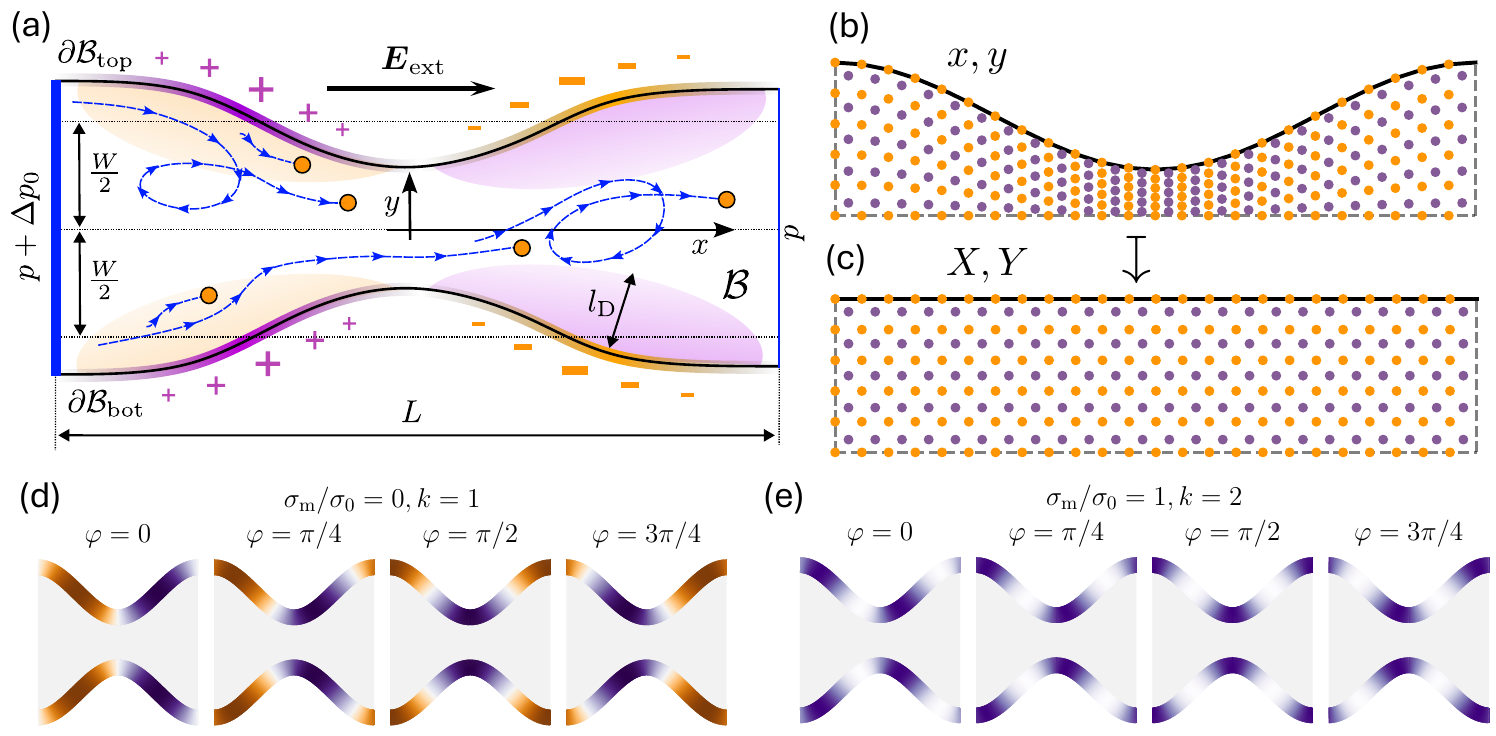}
    \caption{(a) Concept diagram of electrochemical flow through a wavy channel with spatially varying surface charge density. The orange spheres represent Brownian tracer particles advected by the flow. Numerical discretization of the field variables in (b) the physical domain and (c) the transformed domain; grid point resolution is reduced for clarity. In a staggered arrangement, the velocity, electrostatic potential, and ion concentrations are evaluated at the orange points, and the pressure is evaluated at the purple points. Using the problem's symmetry, the governing equations are evaluated numerically in the domain of a half-pore. Bottom panels show representative prescriptions of $\sigma_\mathrm{c}(\bm{x})$ for (d) one oscillation of net-neutral surface charge and (e) two oscillations of charge with $\sigma_\mathrm{m}=\sigma_0$.}
    \label{fig:Fig1}
\end{figure}

\section{Theoretical background and model design}
\label{sec:model}

\subsection{Governing equations and assumptions}
\label{subsec:pdes}

We consider steady-state two-dimensional flow of an electrolyte through a charged nanochannel with variable aperture, with coordinates $\bm{x} = (x,y)$. The domain $\mathcal{B}$, depicted in Fig.~\ref{fig:Fig1}(a), is bounded by sinusoidal top and bottom walls $\partial\mathcal{B}_\mathrm{top}$ and $\partial\mathcal{B}_\mathrm{bot}$. We set the mean width of the channel to $W$, such that the coordinates of the boundaries are defined by
\begin{equation}
\begin{aligned}
y &= \pm\frac{1}{2}\left[W - \delta W \cos\left(\frac{2\pi x}{L}\right) \right]\qquad &&\text{for } -\frac{L}{2}<x<\frac{L}{2},
\end{aligned}
\end{equation}
and $L$ and $\delta W$ represent the wavelength and amplitude of the boundaries' undulations. Throughout our study, we assume periodicity of the fields across the entry and exit of the channel (at $x=-L/2$ and $L/2$).

A periodic surface charge density, $\sigma_\mathrm{c}(x)$, is applied to the boundaries of the wavy channel,
\begin{equation}
\label{eq:surface_charge_distribution}
\begin{aligned}
    &\sigma_\mathrm{c}(x)=\sigma_\mathrm{0}\sin\left(\frac{2\pi k x}{L}+\varphi\right) + \sigma_\mathrm{m} \qquad &&\text{for } \bm{x}\in\{\partial\mathcal{B}_\mathrm{top},\partial\mathcal{B}_\mathrm{bot}\},
\end{aligned}
\end{equation}
where $k$ adjusts the wavelength of the charge patterns relative to the aperture profile. The surface charge density is placed symmetrically across the horizontal channel axis but may be offset (\textit{asymmetric}) with respect to the geometric undulations. Here, $\sigma_0$ is the peak surface charge density, $\sigma_\mathrm{m}$ is the mean surface charge, and $\varphi$ is a phase angle that shifts the charge distribution along the channel axis. In the case where $\varphi=0$, the surface charge varies \textit{anti-symmetrically} with respect to the geometric undulations, and in the case where $\varphi=\pi/2$ the surface charge varies \textit{symmetrically} with respect to the geometric undulations. To guarantee electro-neutrality, we balance the surface charge density with the charge density of the electrolyte, $\rho_\mathrm{c}$:
\begin{equation}
\label{eq:electro-neutral}
\int_\mathcal{B} \rho_\mathrm{c}(\bm{x}) \mathrm{d}V + \int_{\partial \mathcal{B}_\mathrm{top} + \partial \mathcal{B}_\mathrm{bot}}\sigma_\mathrm{c}(\bm{x})\mathrm{d}A=0.
\end{equation}
For clarity, Figs.~\ref{fig:Fig1}(d) and~\ref{fig:Fig1}(e) plot the surface charge distribution relative to the geometric undulations for a few cases of $\varphi$ and differing $\sigma_\mathrm{m}$.

The surface-charge distribution in Eq.~(\ref{eq:surface_charge_distribution}) is prescribed rather than determined from equilibrium surface chemistry. On chemically active surfaces, such as metal oxides and silica, protonation, deprotonation, and ion complexation can make the surface charge depend on pH, electrolyte composition, and the local electrostatic potential~\cite{behrens2001charge}. Site-binding and charge-regulation models describe this coupling using material-specific properties such as surface-site densities, equilibrium constants, and interfacial capacitances~\cite{davis1978surface,trefalt2016charge}. In narrow silica channels, EDL overlap can further regulate the interface, reducing the surface-charge density while increasing the zeta potential~\cite{wang2010electrochemical}. These surface-chemical reactions are not included in the present model. Our objective is instead to determine how a given charge pattern---particularly its amplitude and phase relative to the channel corrugation---controls the coupled flow and ion transport. We note that constant-charge models have been used successfully in nanofluidic energy-conversion studies. \citet{vanderheyden2006electrokinetic} predicted enhanced conversion efficiency under strong EDL overlap for slit channels with constant surface charge, and subsequent experiments confirmed that pressure-driven power generation is most efficient in the overlapping-EDL regime~\cite{vanderheyden2007power}.

The flowrate through the nanochannel is controlled by one of two driving forces that is directed parallel to the channel axis. One driving force is a pressure gradient, $\nabla p_0 = \Delta p_0/L$, and is imposed, under periodicity, by supplying a pressure jump to the fluid as it moves from the channel exit, on the right, to the channel entrance, on the left: $\llbracket p \rrbracket = p(x=-L/2,y) - p(x=L/2,y)=\Delta p_0$. In the absence of electrokinetic effects and under steady-state conditions, the pressure gradient drives the velocity field according to Poiseuille flow. When electrokinetic effects are added, the charge patterns cause the formation of localized regions of counter charge that are not easily displaced from the channel walls if the local electric field is large. These physics of flow are encoded in an electrokinetically modified momentum balance or Stokes equation,
\begin{equation}
\label{eq:momentum_balance}
\mu\nabla^2\bm{v} - \nabla p + \rho_\mathrm{c}\bm{E} = \mathbf{0},
\end{equation}
which is evaluated under the constraint of liquid incompressibility, $\nabla \cdot \bm{v} = 0$. The third term is the electrokinetic drift force,
\begin{equation}
\bm{E} = -\nabla\phi + \bm{E}_\mathrm{ext},
\end{equation}
which depends on the local electric field, $-\nabla \phi$, with $\phi$ denoting the local electrostatic potential generated by the distribution of ions and the surface charge density of the channel, and an externally supplied electric field, $\bm{E}_\mathrm{ext}$. We identify $\bm{E}_\mathrm{ext}$ as the second possible driving force imposed to produce bulk flow through the channel and choose it to be uniform in space and directed along the channel axis, $\bm{E}_\mathrm{ext}=E_\mathrm{ext}\mathbf{e}_x$.

Electrokinetic effects become significant when the Debye screening length,
\begin{equation}\label{eq:Debye_length}
    l_\mathrm{D} = \sqrt{\frac{\varepsilon k_\mathrm{B}T}{2 e_0^2 c_0}},
\end{equation}
which depends on the bulk salt concentration $c_0$ and thermal energy $k_\mathrm{B}T$, extends appreciably into the channel width. Here, $k_\mathrm{B}$ denotes the Boltzmann constant and $T$ is the temperature. As a reference, for a NaCl solution at a concentration of $c_0=0.01\,\mathrm{M}$, dielectric permittivity of $\varepsilon\approx 80 \varepsilon_0$ (where $\varepsilon_0$ is the vacuum permittivity), and temperature $T=300\,\mathrm{K}$, one finds $l_\mathrm{D}\approx3\,\mathrm{nm}$. We limit our scale of observation to the order of magnitude of this Debye length by choosing a mean channel width of $W=5.25$ nm. At this scale, molecular dynamics simulations and nanoscale experiments have consistently shown that the classical no-slip condition breaks down at smooth, hydrophobic walls, where weak fluid--solid interactions allow a finite tangential velocity to develop at the boundary~\cite{bocquet2007flow,bocquet2010nanofluidics}. Together with the prescription of no-flux conditions normal to the channel surfaces and axial periodicity, we express the boundary conditions for the velocity field as
\begin{subequations}
\label{eq:stokes_flow_bcs1}
\begin{align}
& b \nabla \left(\bm{v}\cdot \mathbf{t}\right)\cdot\mathbf{n} = \bm{v}\cdot\mathbf{t}
&\text{for } \bm{x}\in\{\partial\mathcal{B}_\mathrm{top},\partial\mathcal{ B}_\mathrm{bot}\}, \\
\label{eq:no_flow_advection}
& \bm{v}\cdot\mathbf{n} = 0
&\text{for } \bm{x}\in\{\partial\mathcal{ B}_\mathrm{top},\partial\mathcal{ B}_\mathrm{bot}\}, \\
&\bm{v}(x=- L/2) = \bm{v}(x= L/2) &\text{for } -(W+\delta W)/2 < y < (W+\delta W)/2.
\end{align}
\end{subequations}
Above, $b$ measures the slip length and $\mathbf{n}$ ($\mathbf{t}$) is the unit normal (tangent) vector along $\bm{x}\in \{\partial \mathcal{B}_\mathrm{top},\partial \mathcal{B}_\mathrm{bot}\}$ pointing inward toward the channel axis (axially along the channel axis).

We emphasize that the slip boundary condition we impose is not an \textit{electroosmotic} or \textit{Smoluchowski-type} slip phenomenon~\cite{probstein2005physicochemical}, but a \textit{hydrodynamic} or \textit{Navier-type} condition. In general, $b$ depends on the interaction of the fluid with the boundary, \textit{i.e.}, the wettability of the wall and its molecular topology~\cite{bocquet2007flow}. Our chosen value of $b=20$ nm, as listed in Table~\ref{tab:parameter_values}, thus corresponds to a hydrophobic, smooth boundary with a contact angle of $\theta=110^\circ$~\cite{bocquet2010nanofluidics}. Recent linear-response theory for corrugated nanochannels further shows that spatial variations in the Navier slip length can couple strongly to the channel geometry. Both the magnitude and placement of hydrophobic regions modify electroosmotic and pressure-driven transport, with slip localized near a constriction producing particularly strong enhancement~\cite{carusela2025transport}. To isolate the effects of the surface-charge pattern and geometric corrugation from those of wettability patterning, we therefore prescribe a spatially uniform $b$ and keep it fixed across all simulations. The surface-charge amplitude $\sigma_0=0.5\,e_0\,\mathrm{nm}^{-2}\approx0.08\,\mathrm{C\,m}^{-2}$ falls within the range measured for Pyrex and silica nanoslits at moderate-to-high pH~\cite{schoch2005effect,stein2004surface} and is comparable to the amplitude employed by \citet{curk2024discontinuous} for a similarly confined charge-patterned channel. Thus, although $\sigma_0$ is not calculated from a material-specific charge-regulation model, its magnitude is representative of experimentally accessible charged interfaces. We therefore treat $b$ and $\sigma_0$ as independently adjustable parameters --- a combination physically motivated by electrochemically gated carbon nanochannel geometries, in which the intrinsic hydrophobicity of the carbon surface provides large slip~\cite{secchi2016massive} while an externally applied gate voltage independently controls the surface charge amplitude~\cite{karnik2005electrostatic,daiguji2004ion}.

Following \citet{curk2024discontinuous}, we examine the flow regimes that arise under varying relative magnitudes of $\nabla p$ and $\rho_\mathrm{c}\bm{E}$ in Eq.~(\ref{eq:momentum_balance}) by coupling the momentum equation to conservation equations for cation (+) and anion (-) concentrations ($c_\pm$) via the Poisson-Nernst-Planck (PNP) equations. Specifically, the concentration profiles adhere to steady-state mass balance,
\begin{equation}
\label{eq:continuity}
\nabla\cdot\bm{j}_\pm =\nabla\cdot\left(c_\pm \bm{v}_\pm\right) = 0,
\end{equation}
where $\bm{v}_\pm$ are the velocities of each species, which are calculated relative to the velocity of the background flow, $\bm{v}$, using the slip relations $\bm{v}_\pm - \bm{v}= -M_\pm (k_\mathrm{B}T\nabla \ln(c_\pm/c_0)- z_\pm\bm{E})$. Inserting the slip velocities into Eq.~(\ref{eq:continuity}) yields
modified advection-diffusion dynamics that are supplemented by charge-dependent electrostatic forces, 

\begin{equation}\label{eq:continuity_sub}
\bm{v}\cdot \nabla c_\pm =  D_\pm \nabla^2 c_\pm - M_\pm z_\pm \nabla \cdot (c_\pm \bm{E}).
\end{equation}
Above, $D_\mathrm{\pm}=k_\mathrm{B}TM_\pm$ are the species' diffusivities with $M_\pm$ being the species' mobilities, $z_\pm=\pm e_0$ for a monovalent salt, and $c_0$ is the bulk electrolyte concentration. Although $c_0$ does not appear explicitly in Eq.~(\ref{eq:continuity_sub}), it sets the Debye length scale and enters the conservation dynamics through the equilibrium reference state. Eq.(\ref{eq:continuity_sub}) thus functions to couple the diffuse part of the EDL to the advection of the background flow.

As the background flow is controlled by the no-flow condition in Eq.~(\ref{eq:no_flow_advection}), the remaining part of the ion flux, $\bm{j}_\pm$, along the channel boundaries is controlled by balancing molecular diffusion with electrokinetic drift:
\begin{equation}
\label{eq:ion_flux_bc}
\begin{aligned}
\bm{j}_{\pm}\cdot \mathbf{n} &= \left(\nabla c_\pm \mp l_\mathrm{G}^{-1} c_\pm \bm{E}\right)\cdot \mathbf{n}=0 \qquad &&\text{for } \bm{x}\in\{\partial\mathcal{B}_\mathrm{bot},\partial\mathcal{B}_\mathrm{top}\}.
\end{aligned}
\end{equation}
In the expression above, we introduce the Gouy-Chapman length, $l_\mathrm{G}=\varepsilon k_\mathrm{B}T/(e_0 \sigma_0)$, to indicate the distance from the channel boundary at which the thermal energy of the ions approximates their electrostatic potential energy.

To close the system, the electrostatic potential, and thus the local electric field, is solved using the Poisson equation,
\begin{equation}
\label{eq:Poisson}
-\varepsilon \nabla^2 \phi = \rho_\mathrm{c},
\end{equation}
where a uniform dielectric permittivity is assumed for the background fluid, $\varepsilon=\varepsilon_\mathrm{r}\varepsilon_0$, denoting $\varepsilon_\mathrm{r}$ as the relative permittivity. The local charge density $\rho_\mathrm{c}=z_- c_- + z_+ c_+$ is directly related to the ion distributions provided by Eq.~(\ref{eq:continuity}) and influences the background velocity through Eq.~(\ref{eq:momentum_balance}). For the remainder of the paper, we assume the salt to be monovalent, $z_\pm=\pm e_0$.

Lastly, the surface charge distribution introduced in Eq.~(\ref{eq:surface_charge_distribution}) and conceptually illustrated in Fig.~\ref{fig:Fig1} enters the system through the Neumann boundary conditions that are prescribed along the top and bottom surfaces of the channel,
\begin{equation}
\label{eq:BC_surf_charge}
-\varepsilon\nabla\phi\cdot \mathbf{n} = \sigma_\mathrm{c}(\bm{x}) \qquad\qquad\qquad \text{for } \bm{x}\in \{\partial \mathcal{B}_\mathrm{top},\partial \mathcal{B}_\mathrm{bot}\}.
\end{equation}
Neumann conditions are widely used to study electrokinetic transport along nonuniformly charged interfaces~\cite{ajdari1995electro,ajdari1996generation,malgaretti2019driving,curk2024discontinuous,shrestha2025self,shrestha2025universal}. They are also physically relevant to chemically patterned surfaces~\cite{vlassiouk2007nanofluidic}, externally controlled interfaces, and the weak-regulation limit in which the ionizable surface groups remain nearly fully dissociated. We therefore interpret $\sigma_\mathrm{c}(x)$ as a prescribed effective charge distribution rather than as a prediction of the equilibrium charge of a particular reactive material. This choice isolates the effects of charge placement and geometry from the surface chemistry that generates the charge and provides a benchmark for future models that couple charge regulation self-consistently to the local electrolyte state.

The sign of $\sigma_\mathrm{c}$ and its gradient do not influence the velocity field in a \textit{flat} channel driven by small pressure gradients ($\nabla p_0\ll \sigma_0 c_0 e_0/\varepsilon$), provided the salt is symmetric in valence and diffusivity. Instead, $\mathrm{sgn}(\sigma_\mathrm{c})$ assigns the polarity of the EDLs along the charge patches, while the surface charge gradient $\mathrm{d}\sigma_\mathrm{c}/\mathrm{d}x$ governs the axial equilibrium potential gradient $\nabla\phi\cdot\mathbf{e}_x$ and thus the electrostatic resistance to advection.

Conversely, in \textit{corrugated} channels the slope of the boundaries induces interference of laterally spaced EDLs. For large $l_\mathrm{D}$, pressure-driven flow though corrugated channels with alternating polarity surface charge patches is expected to provide different velocity patterns than flow through channels with monopolar patches.  For uniformly charged surfaces, \citet{messinger2010suppression} showed that nanoscale roughness can suppress electro-osmotic flow when excess surface conduction is appreciable, because curvature produces spatially nonuniform tangential electric fields and surface currents. In the present simulations, the effects of curvature on the bulk electrolyte, EDL structure, electric field, and flow are resolved directly in the physical corrugated geometry without a small-curvature approximation. However, $\sigma_\mathrm{c}(x)$ is prescribed independently of the local curvature, and neither a surface-current conservation law nor curvature-dependent surface chemistry is included. Within the present framework, a curvature-induced redistribution of surface charge would enter through changes in the amplitude, shape, or phase of $\sigma_\mathrm{c}(x)$. Our parameter study therefore quantifies the transport sensitivity to such changes, but does not predict the surface-chemical response that produces them.

\begin{table}[t]
\caption{\label{tab:parameter_values}Parameter values and physical constants used in the numerical simulations.}
\begin{ruledtabular}
\begin{tabular}{lll}
\textbf{Parameter} & \textbf{Description} & \textbf{Value} \\
\hline
$W$ & Channel width & 5.25 nm \\
$L$ & Wavelength of geometry and charge pattern & 15.75 nm \\
$k$ & Wavelength of the charge pattern & 1 or 2 \\
$T$ & Temperature & 300.0 K \\
$k_\mathrm{B}$ & Boltzmann constant & $1.381\times10^{-23}$ J$\cdot$K$^{-1}$ \\
$N_\mathrm{A}$ & Avogadro's constant & $6.022\times10^{23}$ mol$^{-1}$ \\
$\epsilon_\mathrm{r}$ & Relative permittivity of water & 78.5 \\
$\epsilon_0$ & Vacuum permittivity & $8.854\times 10^{-12}$ F$\cdot$m$^{-1}$ \\
$e_0$ & Elementary charge & $1.602\times 10^{-19}$ C \\
$\mu$ & Dynamic viscosity & 1.00 cP \\
$D_0$ & Cation/anion diffusion coefficient & $10^{-9}$ m$^2\cdot$s$^{-1}$ \\
$\sigma_0$ & Surface charge density amplitude & 0.5 $e_0\cdot$nm$^{-2}$ \\
$b$ & Slip length & 20.0 nm \\
$c_0$ & Bulk salt concentration & 0.003 M -- 0.500 M \\
\end{tabular}
\end{ruledtabular}
\end{table}

Non-dimensionalizing the governing equations with $c_\pm = c_0 \tilde{c}_\pm$, $\rho_\mathrm{c}=e_0c_0\tilde{\rho}_\mathrm{c}$, $\bm{E}=\sigma_0\tilde{\bm{E}}/\varepsilon$, $\bm{v}=U \tilde{\bm{v}}$, and $p= W\nabla p_0\tilde p$, and setting the characteristic velocity to the diffusive scale $U=D_0/W$ with $D_0=D_+=D_-$, Eqs.~(\ref{eq:continuity_sub}), (\ref{eq:momentum_balance}), and (\ref{eq:Poisson}) become
\begin{subequations}
\label{eq:governing_equations}
\begin{align}
\label{eq:continuity_non_dim}
& \tilde{\bm{v}} \cdot \tilde{\nabla} \tilde{c}_\pm = \tilde{\nabla}^2 \tilde{c}_\pm  \mp  \tilde{l}_\mathrm{G}^{-1}\tilde{\nabla} \cdot\left( \tilde{c}_\pm\tilde{\bm{E}}\right), \\
\label{eq:momentum_balance_non_dim}
&\tilde{\nabla}^2\tilde{\bm{v}} + \mathrm{Ev}^{-1}\left(\tilde{\rho}_\mathrm{c}\tilde{\bm{E}} - \Pi \tilde{\nabla} \tilde{p}\right) = \mathbf{0},\\
\label{eq:incompressibility}
& \tilde{\nabla}\cdot  \bm{\tilde{v}}=0, \text{ and}\\
\label{eq:Poisson_non_dim}
&\tilde{\nabla}^2\tilde{\phi} = -\tilde{\sigma}_0^{-1}\tilde{\rho}_\mathrm{c},
\end{align}
\end{subequations}
noting $\tilde{\sigma}_0=\sigma_0/(c_0 e_0 W)$ as the dimensionless amplitude of the surface charge density oscillations and $\tilde{\bm{E}}=-\tilde{\nabla}\tilde{\phi} + \tilde{\bm{E}}_\mathrm{ext}$. All lengths are non-dimensionalized by $W$, so that the dimensionless Debye length is $\tilde{l}_\mathrm{D}=l_\mathrm{D}/W$; complete EDL overlap across the half-channel occurs at $2\tilde{l}_\mathrm{D}=1$. The two dimensionless groups appearing in Eq.~(\ref{eq:momentum_balance_non_dim}) are an electroviscous number,
\begin{equation}
\mathrm{Ev} = \frac{\mu U \varepsilon}{W^2 c_0 e_0 \sigma_0},
\end{equation}
which compares viscous to electrokinetic forces, and an electrokinetic number,
\begin{equation}\label{eq:PI}
\Pi=\frac{\varepsilon\nabla p_0}{c_0 e_0 \sigma_0},
\end{equation}
which compares the strength of the pressure gradient to the electrostatic drift force. The second group is a focus of our study and governs the transition between electrokinetically restricted channel flow and Poiseuille-like flow. Its importance (and magnitude) can be tuned by modifying the electrolyte composition, channel geometry, and surface charge placement. The second driving force of interest, $\tilde{\bm{E}}_\mathrm{ext}=\varepsilon\bm{E}_\mathrm{ext}/\sigma_0$, acts within the diffuse part of the EDLs and drives electroosmotic (EO) flow; $\bm{E}_\mathrm{ext}$ is non-dimensionalised to compare the external field to the magnitude of the local electrostatic field holding the EDLs to the charge patches.

\subsection{Numerical implementation}

The coupled set of partial differential equations~(\ref{eq:continuity_non_dim})--(\ref{eq:Poisson_non_dim}) is solved using a combination of finite difference and finite volume methods implemented in the Python programming language. To permit the equations to be solved on a rectilinear grid, we employ a domain mapping procedure, $(\tilde{x},\tilde{y})\to(X,Y)$, that transforms the curvilinear boundaries in the physical domain, $\tilde{\bm{x}}=\tilde{x}\mathbf{e}_x+\tilde{y}\mathbf{e}_y$, to rectilinear coordinates in a transformed domain, $\bm{X}=X\mathbf{e}_x+Y\mathbf{e}_y$~\cite{thompson1982boundary}. We employ a staggered (MAC, Marker-and-Cell) grid arrangement~\cite{harlow1965numerical}, in which $\tilde{p}$ is defined at cell centers and $\tilde{\bm{v}}$, $\tilde{\phi}$, and $\tilde{c}_\pm$ are defined on cell faces. This layout avoids spurious pressure modes. The grid point mapping for a representative discretization is shown in Figs.~\ref{fig:Fig1}(b,c), and the transformation is described in Sec.~S1 of the Supplemental Material~\cite{SM}. The solution to Eqs.~(\ref{eq:governing_equations}) is found first by initializing $\tilde{\phi}$ and $\tilde{c}_\pm$ at the Poisson-Boltzmann equilibrium in the absence of flow ($\tilde{\bm{v}}=0$) and then pseudo-time stepping the coupled set of equations toward steady-state upon turning on flow. The algorithms for these two steps are presented in Sec.~S2 of the Supplemental Material~\cite{SM}.

\section{Results and discussion}
\label{sec:analysis_discussion}

The results are organized to build toward the central finding of this work: that the phase offset between the surface-charge pattern and channel geometry enables rectified, diode-like ionic transport under purely pressure-driven conditions. In Sec.~\ref{sec:linear-response-comparison} we validate the nonlinear PNPS solver against linear-response theory and identify the strongly overlapping-EDL regime in which the Debye--H\"uckel approximation fails. In Sec.~\ref{subsec:net-neutral} we examine electric field-driven flow with near net-neutral surface charge to establish how broken charge--geometry symmetry generates axial motion and how the flow changes with forcing strength. Next, in Sec.~\ref{sec:flow-under-forcing}, we compare pressure- and electric field-driven configurations and identify two flow regimes separated by the competition between the applied driving force and the electrostatic forces retaining counterions near the charge patches. Finally, in Sec.~\ref{subsec:ion-fluxes} we show how the charge--geometry phase offset controls pressure-driven ionic current and selectivity, and use random-walk particle tracking to connect the continuum fluxes to species-dependent trapping, displacement, and dispersion.

\subsection{Solver validation and breakdown of the Debye-H\"{u}ckel approximation}
\label{sec:linear-response-comparison}

We begin by establishing two baseline results. Firstly, that our PNPS solver faithfully reproduces the analytical linear response (LR) solution where the Debye-Hückel (DH) approximation is valid, and, secondly, that the DH approximation fails by an order of magnitude in the large-Debye-length, high-surface-charge regime that governs the physics studied throughout this paper. Many previous treatments of charge-patterned nanochannel flow adopt the DH approximation ($|z_\pm \phi|/k_\mathrm{B}T\ll 1$, giving $\nabla^2\phi\approx l_\mathrm{D}^{-2}\phi$)~\cite{ajdari1995electro,ajdari1996generation} or neglect advective coupling in the ionic continuity equations~\cite{shrestha2025self}; the comparison below quantifies where this linearization ceases to be adequate.

For a flat channel or a channel in which the surface charge is placed anti-symmetrically along the corrugations, the symmetry of the driving forces predicts the electric field to produce no net flowrate through the channel. Instead, it is understood that the forces acting in the EDLs along periodically arranged surface charge patches generate recirculating regions whose penetration depth into the channel depends on the Debye length~\cite{ajdari1995electro}. We begin our analysis by providing a comparison of our numerical solution of the non-linear system of Eqs.~(\ref{eq:governing_equations}) to an analytical approximation of the momentum balance~(\ref{eq:momentum_balance}) calculated from the linear response (LR) of a stagnant equilibrium system to the external electric field.

In the absence of advection, the Poisson-Boltzmann equation can be solved analytically for a flat channel with sinusoidally varying $\sigma_\mathrm{c}$ by assuming linearity between $\rho_\mathrm{c}$ and $\phi$ (the DH approximation). The resulting equilibrium electrostatic potential is evaluated to be~\cite{ajdari1995electro}
\begin{equation}\label{eq:equilibrium_potential}
\phi^\mathrm{eq}(x,y) = \frac{\sigma_0}{K \varepsilon}\sin\left(qx\right)\frac{\cosh(Ky)}{\sinh(KW/2)},
\end{equation}
with $K^2=q^2 + l_\mathrm{D}^{-2}$, where $q=2\pi k/L$ and $l_\mathrm{D}$ is measured by Eq.~(\ref{eq:Debye_length}). The corresponding equilibrium ion densities are provided by $c_\pm^\mathrm{eq}/c_0=1 \mp e_0 \phi^\mathrm{eq}/(k_\mathrm{B}T)$. We may re-write Eq.~(\ref{eq:momentum_balance}) as
\begin{equation}
\mu \nabla^2 \bm{v} - \nabla p' = 0,
\end{equation}
where the electrokinetic drift force is absorbed into an augmented expression for the pressure $p' = p - \varepsilon l_\mathrm{D}^{-2} (\phi^{\mathrm{eq}})^2/2$. The ability to write the drift force as the gradient of a scalar field leads to the observation that no body forces are present, $\nabla \times (\mu \nabla^2 \bm{v}-\nabla p')=\mu \nabla^2 \omega=0$. Here, $\bm{\omega}=\nabla \times \bm{v}$ denotes the vorticity of the velocity field, \textit{i.e.} $\bm{\omega}=\omega \mathbf{e}_z$. Thus, at equilibrium, when $E_\mathrm{ext}=0$ and $\nabla p_0=0$, no flow presides and $\bm{v}=0$.

\begin{figure}[t]
    \centering
    \includegraphics[width=\linewidth]{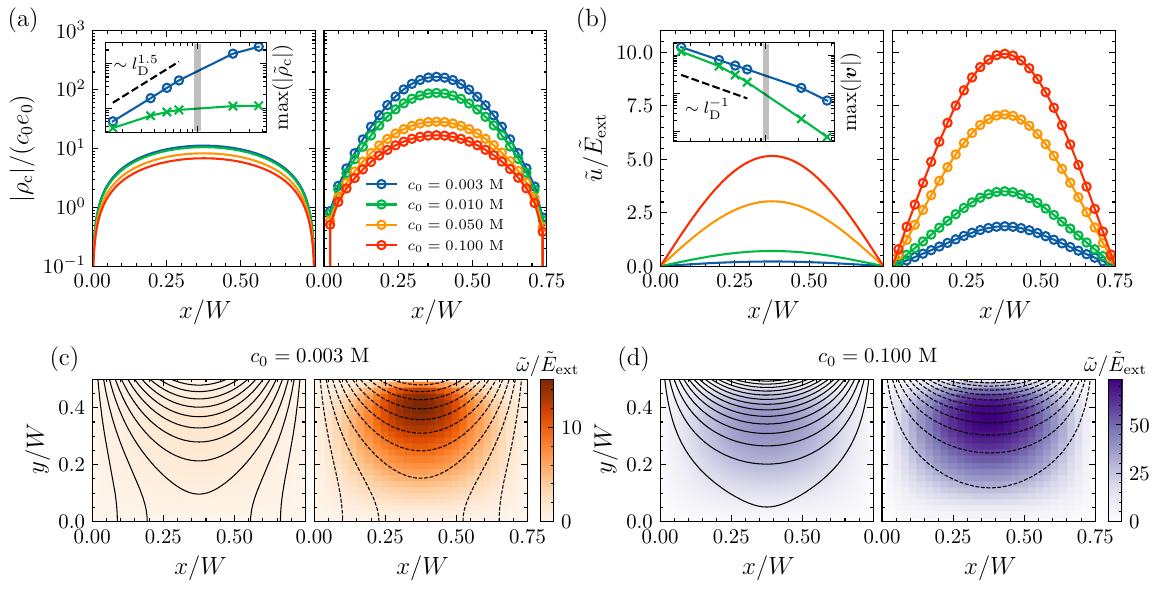}
    \caption{Comparison of linear response theory (LR) with our numerical PNPS solution for electric field-driven flow in a flat channel with net-neutral surface charge using the parameters listed in Table~\ref{tab:parameter_values}; profiles are plotted across a single charge patch $0 \le \tilde{x} \le 0.75$ and the left (right) set of axes plot the LR (numerical) solution. (a) Charge density magnitude $|\tilde{\rho}_\mathrm{c}|$ adjacent to the channel walls at $\tilde{y}=\pm1/2$; the inset shows the channel-maximum of $|\tilde{\rho}_\mathrm{c}|$ as a function of $2\tilde{l}_\mathrm{D}$ for the DH approximation (green) and full PB solution (blue) with the gray vertical line indicating the onset of EDL overlap at $2\tilde{l}_\mathrm{D}=1$. (b) Horizontal velocity magnitude along the channel walls; the inset plots the maximum magnitude of the velocity field probed across the channel. (c,d) Field-normalized vorticity $\tilde{\omega}/\tilde{E}_\mathrm{ext}$ across a single charge patch for (c) $c_0=0.003\,\mathrm{M}$ and (d) $c_0=0.100\,\mathrm{M}$; the LR (left) and numerical (right) solutions are plotted using the same color scale for $\tilde{\omega}/\tilde{E}_\mathrm{ext}$ and are overlaid with iso-potential contours of $\phi^\mathrm{eq}$ at identical values. For all panels, $k=2$.}
    \label{fig:Fig2}
\end{figure}

We perturb the equilibrium solution using an external field. Within the DH approximation, \citet{ajdari1996generation} proposed an analytical solution to Eqs.~(\ref{eq:governing_equations}) by performing the first-order expansion $\bm{E}\approx -\nabla \phi^\mathrm{eq}+E_\mathrm{ext}\mathbf{e}_x$. Choosing $E_\mathrm{ext}\mathbf{e}_x$ as the perturbing field, the Stokes problem for the velocity field (or equivalently the vorticity) is written as
\begin{equation}\label{eq:Stokes-vorticity}
\mu \nabla^2\bm{\omega} - \nabla \rho_\mathrm{c} \times \bm{E} \approx \mu \nabla^2\bm{\omega} - \nabla \delta \rho_\mathrm{c} \times \nabla \phi^\mathrm{eq} + \nabla \rho_\mathrm{c}^\mathrm{eq}\times \bm{E}_\mathrm{ext}=\bm{0},
\end{equation}
and it is further assumed that the distortion of the EDLs' counterion clouds from the equilibrium solution is negligible, $\delta \rho_\mathrm{c}=0$. To provide a comparison with our numerical solution, we adopt \citet{ajdari1996generation}'s approach, modifying the boundary conditions to admit the Navier slip-flow assumptions outlined in Sec.~\ref{sec:model}. The streamfunction derivation used to solve Eq.~(\ref{eq:Stokes-vorticity}) with the boundary conditions in Eq.~(\ref{eq:stokes_flow_bcs1}) is provided in Appendix~\ref{app:LRT-solution}. The resulting velocity profiles take on the form
\begin{subequations}\label{eq:LR-velocities}
\begin{align}\label{eq:LR-velocities-u}
u &= \frac{\sigma_0 E_\mathrm{ext}l_\mathrm{D}^2}{\mu}\sin(qx)\frac{\mathrm{d}g(y)}{\mathrm{d}y};\\
\label{eq:LR-velocities-v}
v &= -\frac{\sigma_0 E_\mathrm{ext} l_\mathrm{D}^2}{\mu}\cos(qx)qg(y),
\end{align}
\end{subequations}
where $w=W/2$. The coefficients imposed by symmetry, no-flow, and slip-flow boundary conditions and the dimensionless function $g(y)$ are derived in Appendix~\ref{app:LRT-solution} and reported in Eqs.~(\ref{eq:LR-coefficients}) and (\ref{eq:LR-g}).

\begin{figure}[t]
    \centering
    \includegraphics[width=\linewidth]{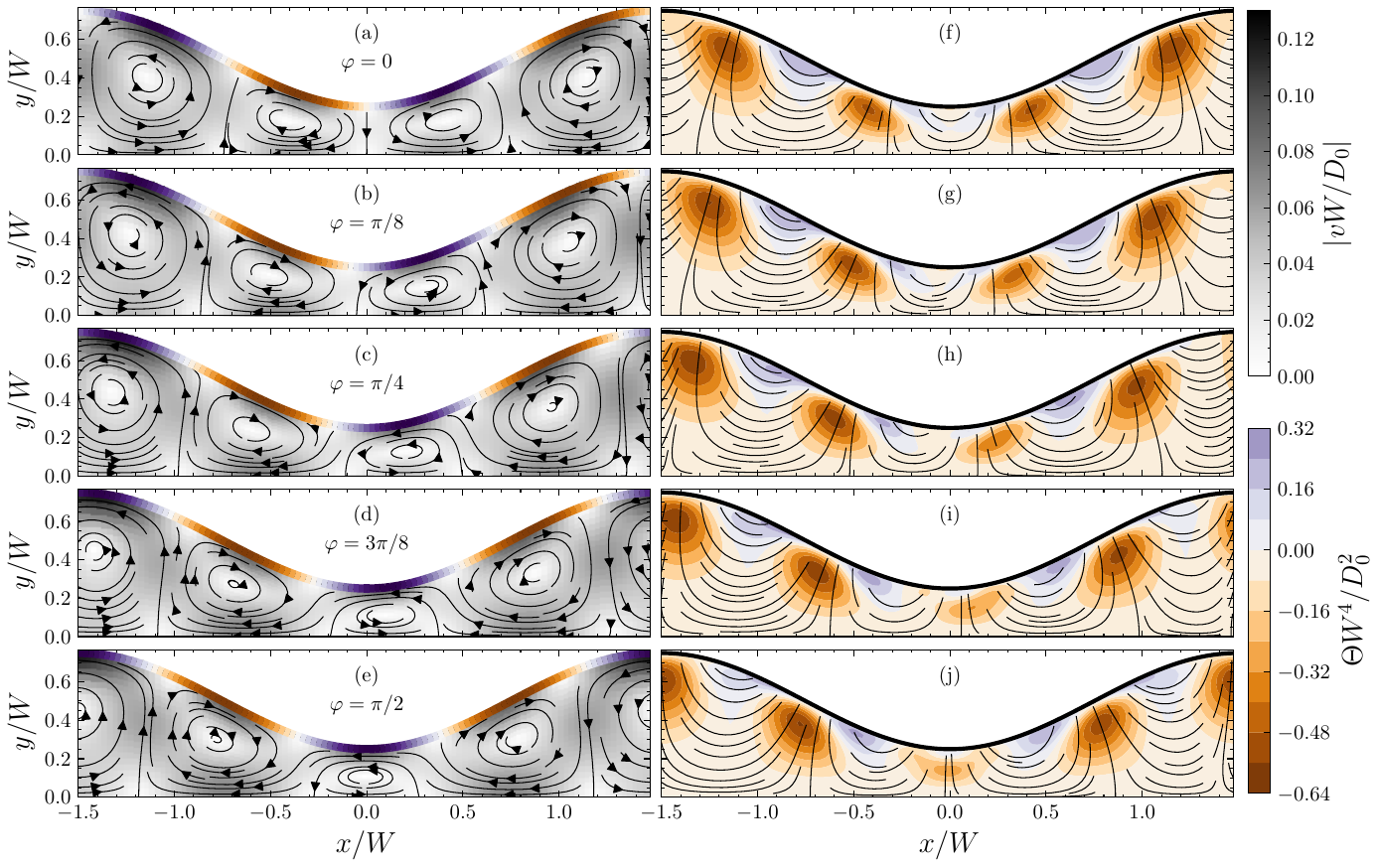}
    \caption{(a-e) Regime I velocity fields induced by an external electric field \textit{below} the flow transition, $\tilde{E}_\mathrm{ext}=1.73\times10^{-2}$, in corrugated channels with $\sigma_\mathrm{m}=0$ and $k=2$ in the absence of an external pressure gradient ($\nabla p_0 = 0$ Pa). Profiles are plotted in terms of the physical coordinates of the top half of the channel across one wavelength of the geometric undulations for varying degrees of asymmetry with the surface charge oscillations. The plotted fields corresponds to $\delta \tilde{W} = 0.5$ and $c_0 = 0.05$ M, with the remaining parameters listed in Table~\ref{tab:parameter_values}. The grayscale colormaps display the magnitude of the velocity profiles with streamlines indicating the direction of flow. The colored line along the top boundary indicates the prescribed surface charge density --- purple indicates positive charge and orange indicates negative charge --- which is shifted along the channel axis moving from (a) to (e). (f-j) Profiles of the normalized Okubo-Weiss parameter, $\tilde{\Theta}$, with streamlines indicating the alignment of the electric field, $\tilde{\bm{E}}$. Positive (negative) values of $\tilde{\Theta}$ identify zones of elongational (rotational) flow.}
    \label{fig:Fig3}
\end{figure}

Figs.~\ref{fig:Fig2}(a) and \ref{fig:Fig2}(b) compare the LR solution expressed in Eqs.~(\ref{eq:equilibrium_potential}) and (\ref{eq:LR-velocities}) to the full non-linear numerical solution for (a) the fluid's charge density and (b) the horizontal velocity profile measured along the boundary of the channel. The data presented is normalized by $E_\mathrm{ext}$, which we chose to be small when performing the numerical simulations to prevent contributions of ion streaming. In panels (a) and (b), the LR and PNPS solutions are plotted side-by-side covering the lateral extent of single surface charge, demonstrating that $\rho_\mathrm{c}$ and $u$ match reasonably well at high salt concentration (when the Debye length and electric potential are small), although they become increasingly dissimilar in shape and magnitude as $c_0$ is decreased. Indeed, the linearized $\rho_\mathrm{c}(\bm{x})$ varies by over an order of magnitude relative to the fully nonlinear Poisson-Boltzmann (PB) solution along the parts of the boundary with the highest $\sigma_\mathrm{c}(\bm{x})$. The insets of the two subplots display the respective maxima of $\rho_\mathrm{c}$ and $|\bm{v}|$ sampled across $\mathcal{B}$. The full nonlinear solution demonstrates power-law scaling for both $\rho_\mathrm{c}$ and $|\bm{v}|$ that persists deep into screening layer overlap, while the LR solution saturates rapidly as $l_\mathrm{D}\simeq W/2$.

The bottom subplots in Fig.~\ref{fig:Fig2} show the low-field-strength vorticity profiles for the LR and numerical solutions across a single surface charge patch for (c) large Debye screening length with $c_0=0.003$ M and (d) shallow Debye screening length with $c_0=0.100$ M. As expected, due to the symmetry of the problem, no directed channel flow is generated. The LR approximation dramatically underestimates the magnitude of the vorticitiy fields, by up to an order of magnitude in the case of $c_0=0.003$ M, though the shapes of the recirculating patterns match excellently between numerical and analytical results. On top of the colormaps, contours indicate the isopotential lines of $\phi^\mathrm{eq}$, which demonstrate excellent agreement between the electric fields calculated using the LR and numerical approaches. The comparison highlights the diminished sensitivity and order-of-magnitude underestimation of both $\rho_\mathrm{c}$ and $|\bm{v}|$ by the LR approximation at low $c_0$ --- precisely the regime of strong EDL overlap and large surface charge that governs the transitions studied in subsequent sections --- and confirms that the full nonlinear PNPS solver is required throughout this paper.

\begin{figure}[t]
    \centering
    \includegraphics[width=\linewidth]{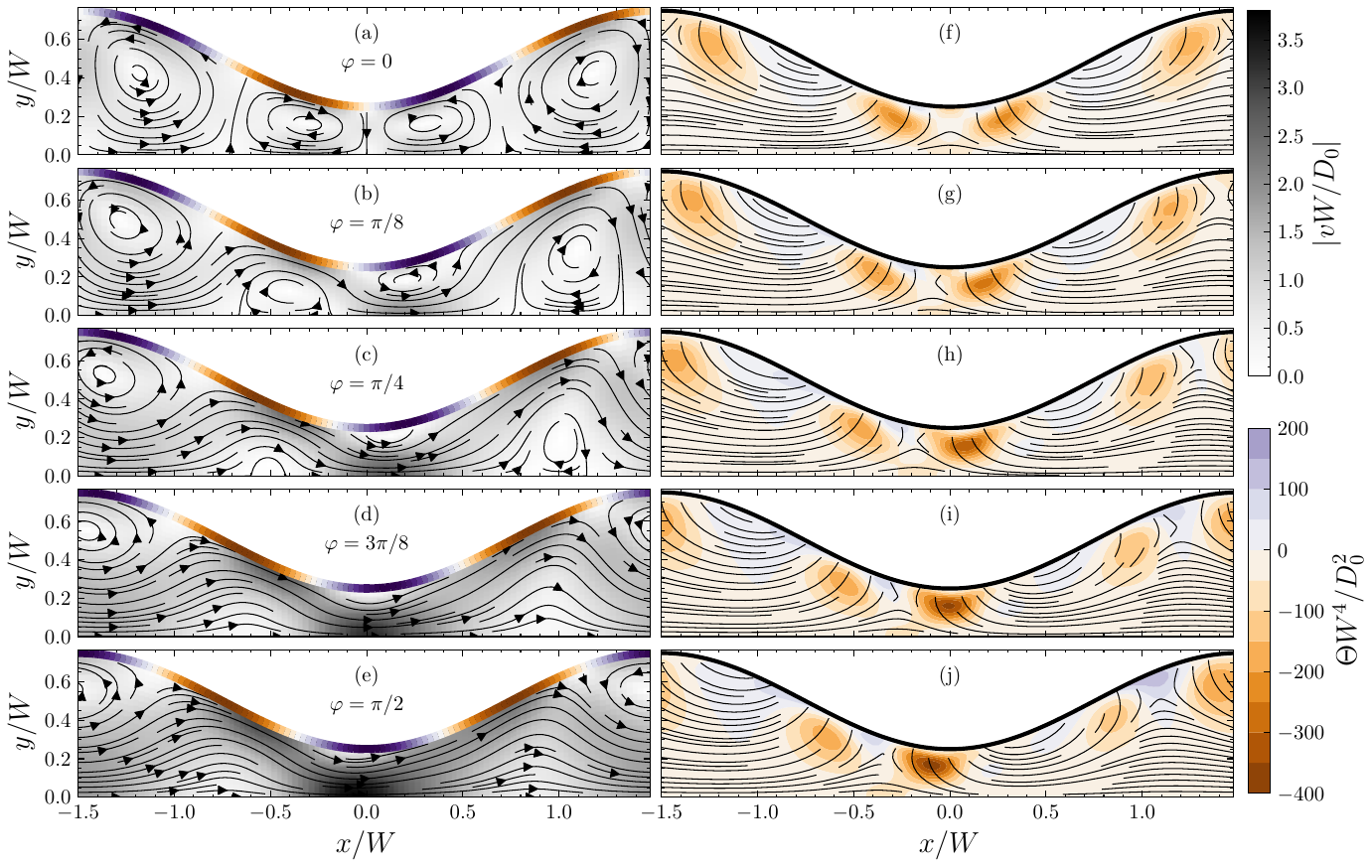}
    \caption{(a-e) Regime II velocity fields induced by an external electric field \textit{above} the flow transition, $\tilde{E}_\mathrm{ext}=4.11\times10^{-1}$, in corrugated channels with net-zero surface charge. All other parameters correspond to the choices made for Fig.~\ref{fig:Fig3}. Similarly, panels (f-j) plot profiles of the normalized Okubo-Weiss parameter, $\tilde{\Theta}$, and the alignment of the electric field.}
    \label{fig:Fig4}
\end{figure}

\subsection{Electric field-driven flow in nanochannels with near net-neutral surface charge}
\label{subsec:net-neutral}

\begin{figure}[t]
    \centering
    \includegraphics[width=\linewidth]{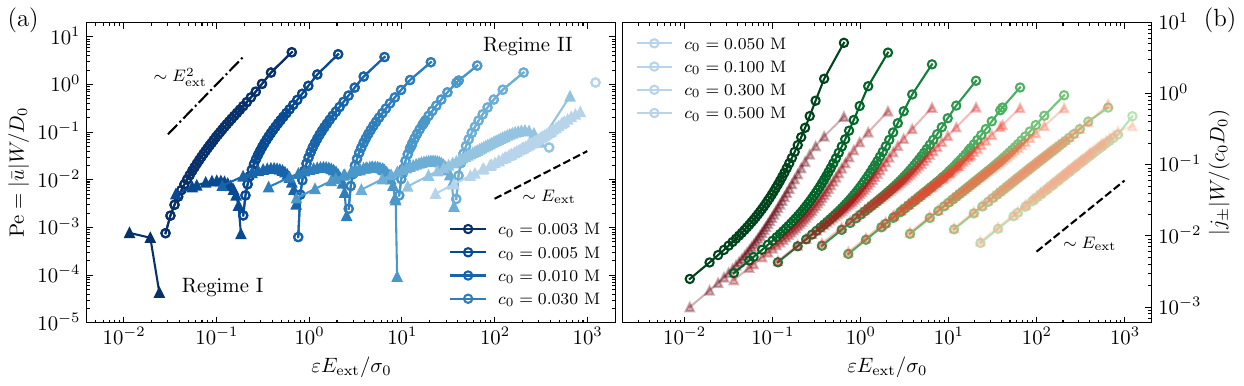}
    \caption{(a) Magnitude of the mean flow rate as a function of the external electric field strength along a channel with $\sigma_\mathrm{m}=0$, $\varphi=\pi/2$, and $k=2$. All curves correspond to $\delta \tilde{W}=0.5$ and the parameter values listed in Table~\ref{tab:parameter_values}; curves for salt concentrations above $c_0=0.003$ M are offset horizontally for clarity. (b) Magnitude of the cationic (green) and anionic (red) fluxes as a function of the electric field strength. For both panels, mean transport recorded in the negative $\mathbf{e}_x$-direction is indicated by $\blacktriangle$ markers and mean transport in the positive $\mathbf{e}_x$-direction is indicated by $\circ$ markers.}
    \label{fig:Fig5}
\end{figure}

Confident in the implementation of our model (see Sec.~\ref{sec:linear-response-comparison}), we examine the transition between linear and nonlinear flow regimes and the effect geometry has on the flowrate and ionic fluxes. To start, we focus on the previously introduced system of sinusoidal surface charge with $\sigma_\mathrm{m}=0$, which is driven by an external electric field. Although $\sigma_\mathrm{m}$ is set to zero in Eq.~(\ref{eq:surface_charge_distribution}), introducing corrugations into the channel renders the arclength of the boundaries non-uniform in $x$ and causes the integrated surface charge to be non-neutral for some choices of $\varphi$. The net surface charge along one of the boundaries for a single wavelength of the geometry measures
\begin{equation}\label{eq:net-charge}
\Sigma_\mathrm{c}=\int_{-\tilde{L}/2}^{\tilde{L}/2}\sigma_\mathrm{c}(\tilde{x})\sqrt{1+\sin^2(\tilde{x})}\mathrm{d}\tilde{x},
\end{equation}
which computes to approximately 0.0\% and 13.3\% of the total charge along the boundaries for $\varphi=0$ and $\varphi=\pi/2$, respectively, when $k=2$. When $k=1$, Eq.~(\ref{eq:net-charge}) provides $\Sigma_\mathrm{c}=0$ for all choices of $\varphi$. Despite the net charge, the principal driving mechanism of axial flow is symmetry breaking rather than a directed net force: the flow reversal observed in Fig.~\ref{fig:Fig5}(a) as $E_\mathrm{ext}$ crosses the Regime I--II transition is inconsistent with net-charge-driven flow, which would produce monotonic, unidirectional transport.

With the system defined, we ask the questions: \textit{(1) To what effect does broken symmetry between surface charge placement and geometry generate axial flow?} And, \textit{(2) How does amplifying the external field above the local electrokinetic drift force that binds counterions to the EDLs alter the character of the flow?} We focus most of our analysis on the case of $k=2$, when the wavelength for the charge oscillations is half that of the corrugation, which showed more interesting nonlinear behavior for electric field-driven flow compared to the case of $k=1$. We designate the flow regime at low driving force as Regime I (i.e., when the diffuse parts of the EDLs maintain close resemblance to the shapes of their stationary equilibrium profiles). The left panels in Fig.~\ref{fig:Fig3} display characteristic velocity profiles in Regime I generated by a low magnitude external electric field in a channel with a Debye length of $2\tilde{l}_\mathrm{D}=0.52$ and geometric undulations of amplitude $\delta \tilde{W}=0.5$. Moving from top to bottom in Fig.~\ref{fig:Fig3}, the surface charge distribution is varied from being (a) anti-symmetric to being (e) symmetric with respect to the geometrical configuration. All cases of the phase angle produce four circulation regions centered over the location of the charge patches. Upon breaking the symmetry of the flow structure (see Figs.~\ref{fig:Fig3}(b-e)), a meandering pathway is generated between the circulation regions that increasingly projects a net flux along the channel axis. The axial flow path becomes more pronounced as $\varphi$ is shifted from $0$ to $\pi/2$, adopting the flow direction set by the surface conduction nearest the channel constriction. Notably, the maximum velocity reaches only a fraction of the molecular diffusion rate for salts moving across the channel.

To visualize and quantify the changes to the kinematics, we compute the Okubo-Weiss metric~\cite{okubo1970horizontal},
\begin{equation}
\Theta(\bm{x}) = -4 \det\left(\dot{\bm{e}}\right),
\end{equation}
where $\dot{\bm{e}} \equiv \nabla \bm{v}$ is the strain rate tensor. The Okubo-Weiss metric is a topological parameter: $\tilde{\Theta}=\Theta W^4/D_0^2 \geq 0$ indicates flow dominated by shear and normal strain, while $\tilde{\Theta}<0$ indicates rotation-dominated flow~\cite{okubo1970horizontal,de2012flow,basilio2022linking}. It thereby identifies zones of intense vorticity, shear, and fluid compression that are relevant to solute residence times and dilution.

We plot $\tilde{\Theta}$ in Fig.~\ref{fig:Fig3}(f-j) along with the electric field lines. Lobes indicating strong rotational deformation are more pronounced along the boundary near the recesses of the channel. The vorticity-dominated regions are suppressed at the channel constriction where the velocity profile is compressed and viscously interferes with the vortices generated along the opposing boundary (i.e., for $\varphi=\pi/2$). Extensional flow is observed in the zones between the charge patches where the flow splits into adjacent recirculating regions. We also note that all electric field lines originate from one charge patch and terminate at an adjacent charge patch of opposite polarity.

As the magnitude of $\bm{E}_\mathrm{ext}$ is increased, the morphology of the velocity profiles changes. At large driving force, the shape of the counterion profiles in the diffuse parts of the EDLs becomes significantly distorted, causing $E_\mathrm{ext}$ to nonlinearly affect the flow profiles. We designate the flow regime at which viscous stresses exceed the local electrokinetic drift force binding the counterions to the charge patches along the boundaries as Regime II. Fig.~\ref{fig:Fig4} plots the $\tilde{v}$- and $\Theta$-fields for Regime II at high $|\bm{E}_\mathrm{ext}|$. Due to symmetry, the case of $\varphi=0$ maintains close semblance to its Regime I velocity profile, though the centers of the circulation currents are pulled inward slightly, in the direction of the EO force. As the symmetry is broken ($\varphi=\pi/8,\,\pi/4,\,3\pi/8,$ and $\pi/2$) the EO forces broaden the axial flow patterns, which increasingly wash away the vortex structures; the vortices in the troughs of the channel are most resistant to displacement. Interestingly, transitioning from Regime I, the net-axial flow path changes direction and concentrates proximally across the channel constriction. These kinematics are detailed in the right panels of Fig.~\ref{fig:Fig4}. The $\Theta$ maps show that as $\varphi\to\pi/2$ rotational flow increasingly concentrates at the constriction, where flow is forced to separate from the boundary and moves in the direction opposite the local EO force; the velocity field is thus driven by the negatively charged (orange) patches at the gradients in corrugation. The velocity achieves magnitudes that are many times that of the characteristic diffusion rate of the ions and can be driven fast by increasing $E_\mathrm{ext}$. We also observe the elongation of the electric field lines, many of which no longer begin and end at adjacent charge patches.

The salt concentration of the electrolyte plays an important role in the momentum balance and the transition from Regime I to Regime II. Figure~S1 in the Supplemental Material~\cite{SM} displays patterns in Regime II at the same $E_\mathrm{ext}$ but differing $l_\mathrm{D}$. Shallow screening layers maintain dominant regions of rotational flow, whereas electrolytes with more diffuse screening layers, extending across the channel width, completely suppress the circulation patterns. The flow regimes are readily identified in Fig.~\ref{fig:Fig5}(a), which plots the P\'{e}clet number, $\mathrm{Pe} = |\bar{u}|W/D_0$, with the longitudinal mean velocity given by
\begin{equation}\label{eq:mean-vel}
\bar{u}=\frac{1}{V}\int{\bm{v}(\bm{x})\cdot\mathbf{e}_x\mathrm{d}V},
\end{equation}
against $E_\mathrm{ext}$. Note that $V$, in Eq.~(\ref{eq:mean-vel}), represents the volume of the channel across a single wavelength of the corrugation, $-1/2<x/L\le1/2$. The brightness of the curves in Fig.~\ref{fig:Fig5} represents the salt concentration of the electrolyte, and each curve is offset horizontally for clarity. Regime I is characterized by a linear relation between $\mathrm{Pe}$ and $E_\mathrm{ext}$ for all $c_0$. As $E_\mathrm{ext}$ is increased, the flowrate for most $c_0$ switches direction and introduces a nonlinear scaling with $E_\mathrm{ext}$ that is enhanced as $c_0$ is decreased. The nonlinear scaling extends well beyond the Regime I-Regime II transition. We attribute this nonlinearity to the preferential release of counterions from the charge patches along the sloped portion of the channel and the channel constriction --- where viscous stresses are largest --- while shielding the counterions in the troughs. Thus, increasing $E_\mathrm{ext}$ both increases the force on the ions and the net charge of the mobile fluid volume. For large $c_0$ (see, $c_0=0.300$ M and $c_0=0.500$ M), the flowrate is linearly related to $E_\mathrm{ext}$ for nearly all of the values tested. Thus, the surface charge structure and geometry present a gating mechanism for electric field-driven nanochannel flow.

Fig.~\ref{fig:Fig5}(b) presents companion curves to the background flow depicted in Fig.~\ref{fig:Fig5}(a) for the dimensionless fluxes of cations and anions,
\begin{equation}
j_{\pm} = \frac{1}{V}\int_\mathcal{B} c_\pm \bm{v}_\pm\cdot \mathbf{e}_x\mathrm{d}V.
\end{equation}
It is worth pointing out that at low $E_\mathrm{ext}$ most curves show similar cation and anion fluxes (when inspected on a log scale) despite the discrepancy in the integrated surface charge alluded to above. At low $E_\mathrm{ext}$ and small $l_\mathrm{D}$ flux is generated predominantly from ionic conduction in the proximal portion of the channel, where the electrolyte is neutral. When $l_\mathrm{D}\simeq W/2$ and the screening length extends into the channel width, Fig.~\ref{fig:Fig5}(b) shows that the cationic and anionic fluxes differ. Similarly, as flow enters Regime II, the magnitudes of the cationic and anionic fluxes separate due to an unequal suspension of the respective counterions from the surface charge patches and an increasingly unidirectional velocity field. The separation of $j_+$ and $j_-$ tracks the degree of nonlinearity in $\mathrm{Pe}$.

Matching the charge and geometric wavelengths ($k=1$) reduces destructive interference between neighboring recirculation cells and increases the flow rate relative to $k=2$ (Fig.~S2 in the Supplemental Material~\cite{SM}). The antisymmetric alignment ($\varphi=0$) produces two recirculation zones but no net axial flow [Fig.~S2(a)], whereas the symmetric alignment ($\varphi=\pi/2$) generates axial transport while retaining small vortices in the channel troughs [Fig.~S2(b)]. Figure~S2(c) shows the corresponding mean flow rate and ionic fluxes across the transition for low- and high-salt electrolytes. Unlike the case of $k=2$, the $k=1$ configuration produces flowrates that are nearly linear across the full range of $E_\mathrm{ext}$ and the flowrates, in general, exceed the magnitudes of those plotted in Fig.~\ref{fig:Fig5} where smaller $l_\mathrm{D}$ lead to larger velocities. The ionic fluxes exhibit a more pronounced species-dependent response: Viscous drag mobilizes the anions placed along the channel's constriction, while counterions attached to the channel's trough remain trapped even at relatively high flowrates.

\subsection{Flow regimes under directed external forcing}
\label{sec:flow-under-forcing}

In Sec.~\ref{subsec:net-neutral} we explored flow and ion transport for scenarios in which, on average, the fluid volume was subjected to no (or a minimal) directed axial force; axial flow was generated predominantly by recirculating zones that pressed against the wall geometry. In the current section, we outline the ability to control flow using either a pressure gradient, $\Pi$, that drags the fluid across near-neutral surface charge oscillations, choosing $\sigma_\mathrm{m}=0$ as before, or an external electric field, $E_\mathrm{ext}$, with surface charge patches of single-signed polarity, prescribing $\sigma_\mathrm{m}=\sigma_0$.

\begin{figure}[p]
    \centering
    \includegraphics[width=0.8\textwidth]{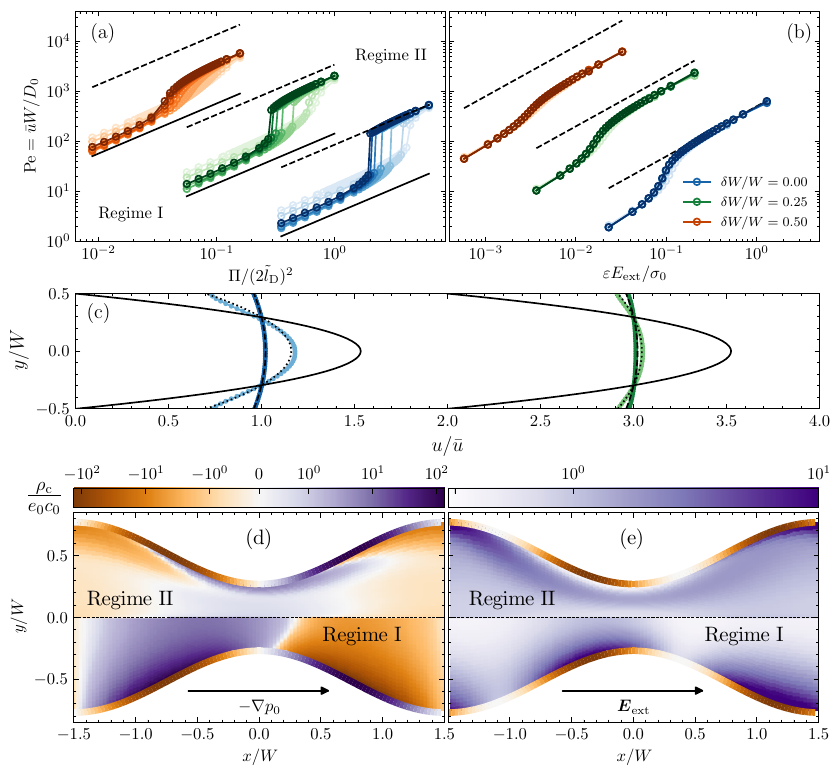}\par\vspace{-1pt}
    \caption{(a) Mean flow rate versus rescaled pressure gradient for $\delta\tilde{W}=0.00$, $0.25$, and $0.50$ (blue, green, and orange), $c_0=0.003$--$0.500\,\mathrm{M}$ (dark to light), and $k=1$; dashed and solid lines denote slip and no-slip Poiseuille flow. (b) Mean flow rate versus external electric field for $\sigma_\mathrm{m}=\sigma_0=-0.25\,e_0\cdot\mathrm{nm}^{-2}$ and $k=2$; the dashed line is the uniformly charged, flat-channel EO limit. Curves in (a,b) are offset diagonally. (c) Mean velocity profiles at $c_0=0.010\,\mathrm{M}$ in Regimes I and II (light and dark) for $\Pi$- and $E_\mathrm{ext}$-driven flow in flat nanochannels (blue and green, respectively; the latter offset by 2.0). Black lines show analytical approximations for flow with no-slip (solid curves) and slip (dashed curves) boundaries. Dotted curves show analytical approximations for the velocity profile with reduced slip lengths of $0.1b$ and $0.4b$ for $\Pi$- and $E_\mathrm{ext}$-driven flow, respectively. (d,e) Charge density distribution in a corrugated channel for $c_0=0.010\,\mathrm{M}$ under (d) pressure and (e) electric-field forcing. Bottom and top halves show Regimes I and II: $\Pi/(2\tilde{l}_\mathrm{D})^2=7.08\times10^{-1}$ and $2.25\times10^{0}$ in (d), and $\tilde{E}_\mathrm{ext}=2.31\times10^{-2}$ and $1.30\times10^{0}$ in (e). Flow moves from left to right; wall colors indicate the surface charge distribution.}
    \label{fig:Fig6}
\end{figure}

We introduce the rescaled pressure gradient $\Pi/(2\tilde{l}_\mathrm{D})^2$, which collapses the salt-concentration dependence of the Regime I--Regime II transition onto a single threshold at $\Pi/(2\tilde{l}_\mathrm{D})^2=\mathcal{O}(1)$ for all $c_0$. The rescaling follows from apportioning the electrokinetic drift force in Eq.~(\ref{eq:PI}) by the EDL volume fraction: reasoning that each charged patch occupies a pore volume $(L/2)(W/2)\simeq(W/2)^2$ and that the EDL covers approximately $l_\mathrm{D}^2$ of this volume gives the factor $(2l_\mathrm{D}/W)^2=(2\tilde{l}_\mathrm{D})^2$. The resulting collapse is confirmed in Fig.~\ref{fig:Fig6}(a), which plots Pe against $\Pi/(2\tilde{l}_\mathrm{D})^2$ for several $c_0$ and corrugation amplitudes $\delta \tilde{W}=0.00$, $0.25$, and $0.50$. Note that curves are plotted for several $c_0$ and differing values for the amplitude of the geometric undulations, $\delta \tilde{W}=0.00$, $0.25$, or $0.50$; for all cases in Fig.~\ref{fig:Fig6}(a) $k=1$ and curves for differing $\delta \tilde{W}$ are offset from one another in the figure for clarity.
We explored a range of phase angles, $\varphi$, for $\sigma_\mathrm{c}(x)$, but its influence on the $\Pi$--$\mathrm{Pe}$ relation was negligible at the scales shown; accordingly, Fig.~\ref{fig:Fig6}(a) presents results for $\varphi=0$ only. However, as discussed in Sec.~\ref{subsec:ion-fluxes}, $\varphi$ strongly controls the ion fluxes.

Introducing a pressure gradient along $\mathbf{e}_x$ clearly evinces two distinct regimes of flow. In Regime I, at low $\Pi$, the mean flow curves exhibit linear scaling with respect to the applied driving force, whose magnitude remains lower than that predicted for Poiseuille flow with the appropriated slip condition through a channel of equivalent mean width: $\bar{u}_\mathrm{P}=-(\nabla p_0/\mu)(W^2 / 24 + b W/4)$. This discrepancy arises from two main factors: (i) the geometric asperities, on average, impede flow more than they enhance it; and (ii) the surface charge patches exert an electrostatic counterforce on the fluid; the counterforce comes from localized streaming potentials that inhibits the removal of ions from the EDLs. The effect of (i) is removed when inspecting flow in a flat channel (blue curves). Notably, all values of $\mathrm{Pe}$ have a lower bound that is well estimated by Poiseuille flow in the limiting case of no-slip, $b\to0$.

\begin{figure}[t]
    \centering
    \includegraphics[width=\linewidth]{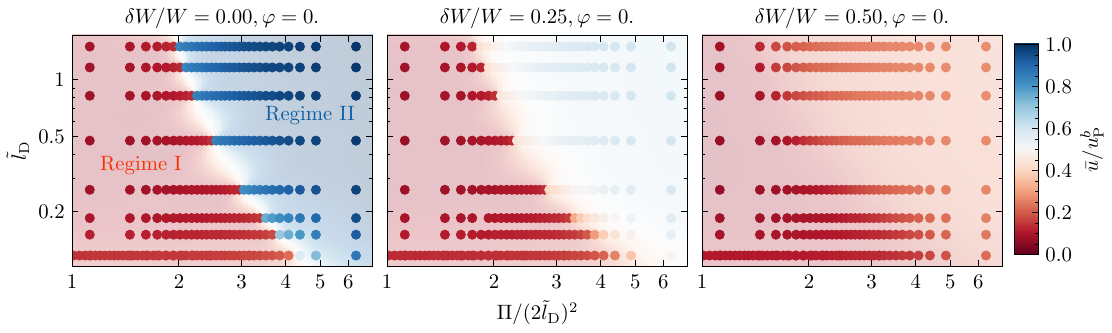}
    \caption{Phase diagram of pressure gradient driven channel flow as a function of the imposed pressure gradient (horizontal axis) and the Debye length (vertical axis) using the parameters listed in Table~\ref{tab:parameter_values}. The colormap displays the mean channel velocity relative to the mean velocity of Poiseuille flow in an uncharged, flat channel; bright dots indicate the locations where data was collected and the muted coloring in between is interpolated. The boundary between Regime I and Regime II becomes increasingly diffuse as the channel transitions from being flat (left) to moderately undulated (center) to highly undulated (right).}
    \label{fig:Fig7}
\end{figure}

As $\Pi$ is further increased (entering Regime II) we observe a transition across which the electrostatic drift force becomes secondary to mechanical pressure --- a two-regime structure consistent with the surface-charge-governed conductance behavior identified in nanoslit experiments~\cite{stein2004surface,schoch2005effect}; the transition and the two flow regimes are identified in finer detail in the phase maps displayed in Fig.~\ref{fig:Fig7}. The phase maps normalize the mean channel longitudinal velocity by the Poiseuille result (with slip boundary conditions) for a flat channel, namely $\bar{u}_\mathrm{P}$ and demonstrate the $c_0$, $\Pi$, and $\delta W$ dependence of the flow. In the case of uniformly placed surface patches, \citet{curk2024discontinuous} showed the transition in flat channels to be discontinuous when $l_\mathrm{D}\approx W/2$: An incremental change in the driving force leads to an orders-of-magnitude change in $\bar{u}$. Figures~\ref{fig:Fig6}(a) and \ref{fig:Fig7} reproduce this transition for sinusoidal surface charge placement and show that the gating mechanism becomes more muted and the transitions are smoothened as the amplitude of the geometric undulations is increased. Similarly, decreasing $l_\mathrm{D}$ localizes the electrokinetic transport resistance to the boundaries and gradually smoothens the gating effect.

As a second scenario, axial flow control is inspected as a function of the external electric field. To introduce a similar number and amplitude of charge patches as imposed on the $\Pi$-driven flow, we set $\sigma_\mathrm{m}=\sigma_0=-0.25\,e_0\,\mathrm{nm}^{-2}$ (the sign of the surface charge is chosen to set the flow direction from left to right) for the $E_\mathrm{ext}$-driven flow; to ensure equivalent surface charge gradients, $|\partial_x\sigma_\mathrm{c}(x)|$, we set $k=2$. With this choice of parameters, the surface charge is purely negative, such that $E_\mathrm{ext}$ generates a strongly directed net force on the fluid acting in the diffuse parts of the EDLs. Sample distributions for $\sigma_\mathrm{c}(x)$ for $\varphi=0$ are shown for the $\Pi$- and $E_\mathrm{ext}$-driven cases along the channel boundaries in Fig.~\ref{fig:Fig6}(d,e). Although $E_\mathrm{ext}$-driven flow also produces distinct flow regimes (see Fig.~\ref{fig:Fig6}(b)), the electrokinetic resistance of the surface charge gradients is softer, leading to a more gradual transition. This observation, made for sinusoidal $\sigma_\mathrm{c}(x)$, differs from the observation by \citet{curk2024discontinuous}, who produced a discontinuous transition for closely spaced, uniformly charged patches. The distinction in behavior of the $\Pi$-driven and $E_\mathrm{ext}$-driven flow may be rationalized as follows: The mechanical driving force operates across the whole cross-section of the nanochannel and pushes counterions of a given charge into adjacent screening clouds of opposite sign. Conversely, the electrochemical driving force localizes in the screening layer and pushes counterions across surface charge patches of similar sign~\cite{ajdari1995electro,mortensen2005electrohydrodynamics}.

As with pressure-driven flow, for sufficiently large $E_\mathrm{ext}$ the counterions are swept from the channel boundaries and mix horizontally. Example charge density profiles in Regime I and Regime II for both $\Pi$- and $E_\mathrm{ext}$-driven flow in a corrugated geometry are provided in Fig.~\ref{fig:Fig6}(d,e). To help visualize the distribution of the charge density near crossovers in sign and in regions proximal to the pore axis, we plot $\tilde{\rho}_\mathrm{c}$ on a linear scale between $-1\le\tilde{\rho}_\mathrm{c}\le1$ and on a signed logarithmic axis for $|\tilde{\rho}_c|>1$. $\rho_\mathrm{c}(\bm{x})$ is significantly more homogeneous in the proximal regions of the channel after entering Regime II. With increasing $\Pi$ or $E_\mathrm{ext}$, the widths of the EDLs reduce as increasingly distal segments of the counterion clouds are removed; this opens up more of the pore volume to advection. For $E_\mathrm{ext}$-driven flow, the stripping of ions from the charge patches leads to an increasingly uniform electrolyte charge density along the boundary.

As a reference case, we approximate the EO velocity profile for a flat channel of uniform surface charge. The solution to the linearized Poisson-Boltzmann equation takes the form $\phi(y)=-\varepsilon^{-1}\sigma_0l_\mathrm{D}\cosh(y/l_\mathrm{D})/\sinh(W/(2l_\mathrm{D}))$. Assuming the charge density profile to remain quiescent, this expression can be inserted into the Stokes equation $\mu \partial_{yy}u=\varepsilon l_\mathrm{D}^{-2}E_\mathrm{ext}\phi(y)$ to approximate EO flow with slip boundary condition as
\begin{equation}\label{eq:EO-flow}
u_\mathrm{EO}(y) = -\frac{\sigma_\mathrm{m} l_\mathrm{D}E_\mathrm{ext}}{\mu\tanh\left(W/(2l_\mathrm{D})\right)} \left(1-\frac{\cosh(y/l_\mathrm{D})}{\cosh(W/(2l_\mathrm{D}))}\right)-\frac{\sigma_\mathrm{m} b E_\mathrm{ext}}{\mu}.
\end{equation}
Fig.~\ref{fig:Fig6}(b) shows that the flowrates, $\bar{u}$, for the blue curves approach the values predicted by cross-sectionally integrating Eq.~(\ref{eq:EO-flow}); notably, curves for different $c_0$ closely overlap in panel (b), reflecting the weak concentration dependence of EO flow in this parameter regime. As before, $\delta W>0$ tends to inhibit flowrates relative to the flat channel counterpart.

The drop in viscous stress along the boundary in moving from Regime I to Regime II can be discerned from the mean horizontal velocity profiles plotted for a flat channel in Fig.~\ref{fig:Fig6}(c). In Regime II, above the transition, the profiles perfectly match the Poiseuille result for the case of $\Pi$-driven flow or the uniform EO flow result for the case of $E_\mathrm{ext}$-driven flow when using our prescribed value for $b$. This demonstrates that the flow does not experience appreciable electrokinetic resistance due to the surface charge gradients. In Regime I, the cross-sectionally averaged velocity profile produces stronger gradients in $u(y)$, causing $u(y)$ to ``bend'' toward the conformation of no-slip Poiseuille or EO flow. Indeed, choosing a slip length of $0.1 b$ for Poiseuille flow or $0.4b$ for EO flow permits close approximation of the horizontally averaged velocity profiles. Hence, the reduction of the equivalent slip length is another approach to quantifying the effect of the charge patches on flow. This mimics the interpretation of \citet{ghosal2002lubrication}, who studied a similar setup of electrochemical transport in the lubrication approximation and showed that the resistance to flow through channels and straight capillaries with axially varying cross-section and zeta potential can be estimated by assigning an equivalent radius and wall charge.

\subsection{Ionic fluxes across pressure-driven flow regimes}
\label{subsec:ion-fluxes}

\begin{figure}[t]
    \centering
    \includegraphics[width=\linewidth]{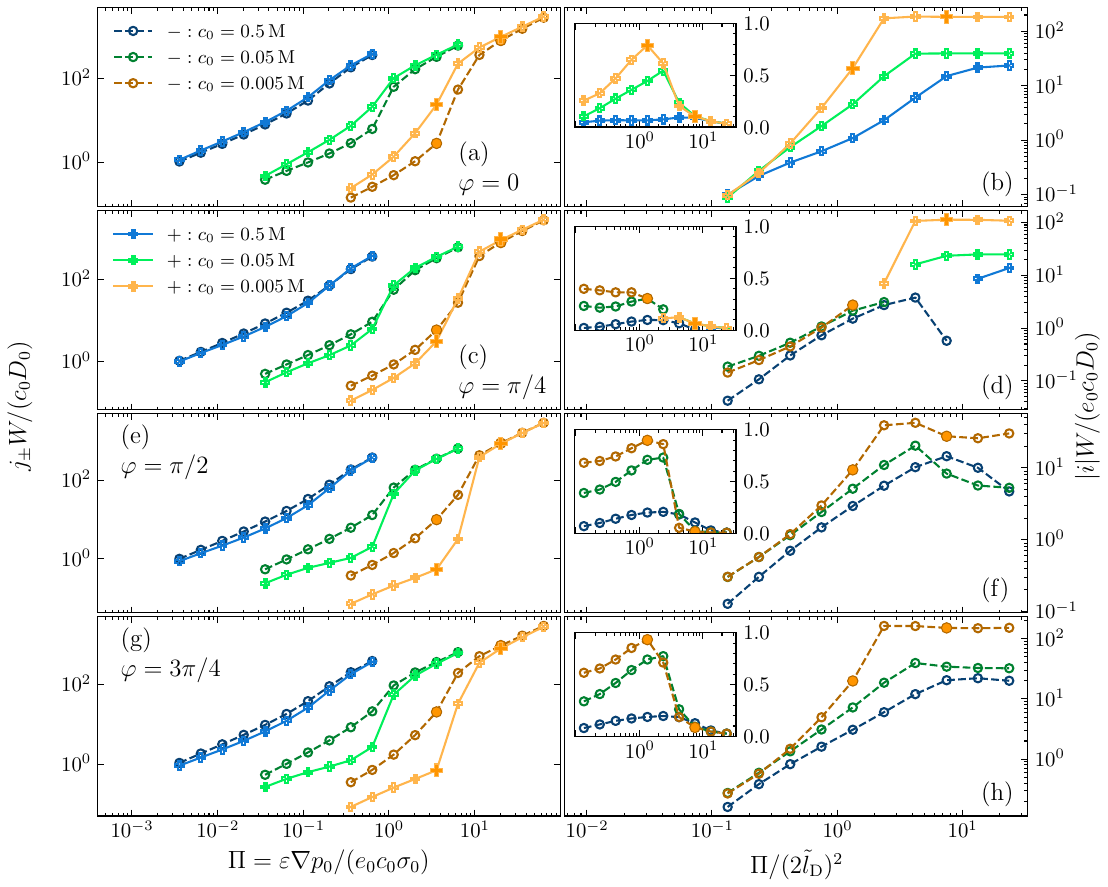}
    \caption{(a,c,e,g) Anion and cation fluxes as a function of applied pressure for $\delta \tilde{W}=0.5$, $k=1$, and the parameter values listed in Table~\ref{tab:parameter_values}. (b,d,f,h) Ionic currents plotted against the rescaled pressure gradients $\Pi/(2 \tilde{l}_\mathrm{D})^2$. The insets plot the selectivity in the ion flux, $|\varsigma|$, against $\Pi/(2 \tilde{l}_\mathrm{D})^2$. Light colored plus symbols indicate positively-charged current and dark colored circles indicate negatively-charged current; filled markers in the left panels correspond to the selected PNPS solutions used as input in the RWPT simulations below. Results are shown for anti-symmetric (a,b; $\varphi=0$), asymmetric (c,d; $\varphi=\pi/4$), symmetric (e,f; $\varphi = \pi/2$), and asymmetric (g,h; $\varphi=3\pi/4$) surface charge alignments with respect to the geometric undulations.}
    \label{fig:Fig8}
\end{figure}

The central finding of this work emerges from the ionic flux analysis: by choosing the phase offset $\varphi$ between the surface charge distribution and the geometric undulations, one can rectify the ionic current in a diode-like fashion under purely pressure-driven flow, selectively activating the passage of cations or anions depending on the direction of the applied pressure gradient. This rectification arises because $\varphi$ controls the positioning of counterion clouds relative to the channel constrictions: when the surface charge maximum is aligned such that the counterions of one species preferentially occupy the high-advection zone at the throat, pressure-driven flow strips that species from the EDL first, generating an asymmetric flux. The influence of $\varphi$ on ionic transport is apparent in the visible separation of $j_+$ and $j_-$ across the transition from electrokinetically restricted to pressure gradient-dominated flow, shown in the left panels of Fig.~\ref{fig:Fig8}, which presents the dimensionless cationic and anionic fluxes as companion curves to the background flow in Fig.~\ref{fig:Fig6}(a). To further characterize the co- or counter-transport of anions and cations we introduce the salt flux,
\begin{equation}\label{eq:salt_flux}
j_\mathrm{s} = \frac{1}{2}\left(j_+ + j_-\right)
\end{equation}
and the ionic current, defined positive for net charge flux directed from left to right,
\begin{equation}\label{eq:ionic_current}
i = z_+j_+ + z_-j_-.
\end{equation}
The selectivity of the ionic flux for our monovalent salt is then calculated by
\begin{equation}\label{eq:selectivity}
\varsigma=\frac{i/e_0}{2j_\mathrm{s}}=\frac{j_+ - j_-}{j_+ + j_-},
\end{equation}
where $-1\le\varsigma\le 1$ and values approaching $-1$ or $1$ indicate channels that are perfectly anion- or cation-selective, respectively.

The right panels in Fig.~\ref{fig:Fig8} plot the ionic current for different choices of $\varphi$. For the case of symmetric surface charge placement ($\varphi=\pi/2$; Fig.~\ref{fig:Fig8}(f)), the boundaries of the narrow part of the channel are positively charged, which generates a negative current for transitional values of $\Pi$; along the constrictions, the counterion clouds obstruct the passage of coions. In approaching the transition, moving toward values of $\Pi/(2\tilde{l}_\mathrm{D})^2\sim 1$, the concentration-normalized ionic current is initially similar for all screening lengths until it begins to fan out as one of the charged species gains favor in passing through the narrowing. This simultaneously enhances the selectivity of the current, shown in the inset of Fig.~\ref{fig:Fig8}(f), with the most effective filtering achieved for large $\tilde{l}_\mathrm{D}$, reaching $|\varsigma|\gtrapprox 0.9$. As $\Pi/(2\tilde{l}_\mathrm{D})^2$ is further increased into Regime II, the selectivity of the ionic current rapidly decreases. We rationalize this as follows: In the pressure-gradient-dominated regime, friction along the boundaries first strips counter-ions from the constrictions and eventually, when flow rates become large enough, also from the troughs, achieving plug-like flow of the electrolyte. We note that the dynamics for $\varphi=\pi/2$ are symmetric with respect to the sign of $\Pi$.

When surface charge is placed anti-symmetrically ($\varphi=0$; Fig.~\ref{fig:Fig8}(b)), pressure-gradient-driven flow provides diode-like behavior for ion transport. That is, when $\Pi>0$ the background flow drives a positive ionic current, $i>0$ (a net flux of positive charge from left to right), and when $\Pi<0$ the background flow also drives a positive ionic current, $i>0$ (a net flux of negative charge from right to left). This behavior is consistent across the transition from Regime I to Regime II as observed by the unipolar current plotted in Fig.~\ref{fig:Fig8}(b); symmetry establishes that cations pass readily for $\Pi>0$ and anions pass readily through channel constrictions for $\Pi<0$. Across the transition the selectivity of the ionic current varies drastically though attains a maximum of $|\varsigma|\approx0.8$ when $\Pi/(2\tilde{l}_\mathrm{D})^2=\mathcal{O}(1)$ and the screening length is large ($c_0=0.005\,\mathrm{M}$; $2\tilde{l}_\mathrm{D}=1.64$), see the inset of Fig.~\ref{fig:Fig8}(b). Selectivity is lost as the pressure gradient is raised into Regime II and most ions advect along the flow. Interestingly, for all $\varphi$, $i$ appears to plateau at high $\Pi$ for most modelled salt concentrations, suggesting that the undulations act as flux barriers and that the character of the transport does not change significantly above the transition pressure.

\begin{figure}[t]
    \centering
    \includegraphics[width=\linewidth]{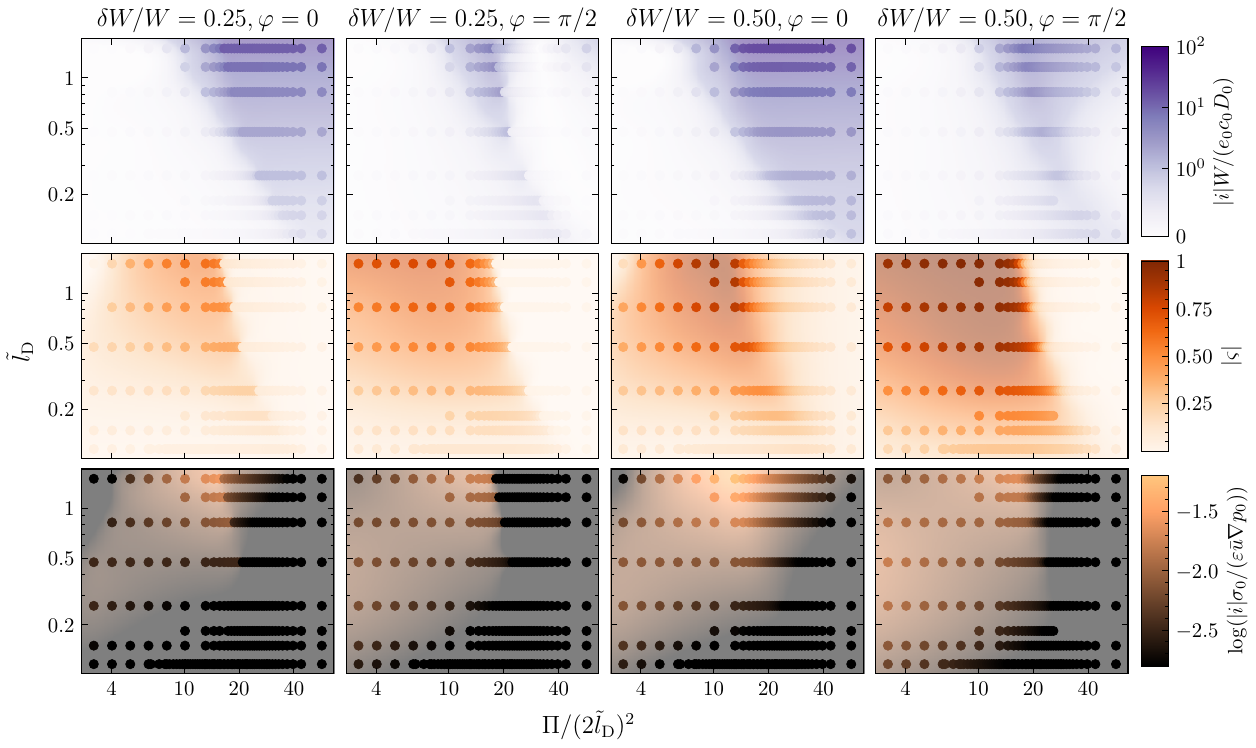}
    \caption{Color maps of the magnitude of the ionic current (top row), selectivity of the ionic flux (center row), and power normalized ionic current (bottom row) for pressure-gradient driven flow. The horizontal axis indicates the rescaled pressure gradient and the vertical axis indicates the non-dimensional Debye screening length. Columns correspond to different choices of $\delta \tilde{W}$ and $\varphi$, as indicated along the top of the figure.}
    \label{fig:Fig9}
\end{figure}

Figs.~\ref{fig:Fig8}(d) and (h) plot the ionic current and selectivity for intermediate phase offsets ($\varphi=\pi/4$, $3\pi/4$); the flux for $\varphi=\pi/4$ would equal the flux for $\varphi=3\pi/4$, measured in the opposite direction, if the pressure gradient was reversed. When the surface charge maximum is placed near the crest and offset to oppose the background advection ($\varphi=3\pi/4$), the selectivity of the current achieves similar or higher values to the antisymmetric case ($\varphi=0$) --- indeed, for $c_0=0.005$ M, the flow achieves near perfect selectivity. This marginal increase in filtering in the forward direction, however, is met with abated filtering when flow is driven in the reverse direction (see the curves for $\varphi=\pi/4$ in Fig.~\ref{fig:Fig8}(d)). Unlike $\varphi=0$, though, the filtering now acts on species of opposite charge, which restores diode-like behavior. At high pressure-gradients, as $\Pi$ is pushed into Regime II, the flow strips ions from the patches of surface charge, first disrupting the structure of the EDLs nearer the channel constrictions and eventually also pulling counterions out of the channel troughs.

To summarize the ion transport behavior, colormaps of the ionic current (top panels) and selectivity (center panels) are resolved over a square parameter space for $\tilde{l}_\mathrm{D}$ and $\Pi/(2 \tilde{l}_\mathrm{D})^2$ for $\varphi\in\{0, \pi/2\}$ and $\delta \tilde{W}\in\{0.25, 0.50\}$; we remind the reader of our previous discussion that $\Pi/(2 \tilde{l}_\mathrm{D})^2$ is independent of $c_0$. Lastly, the bottom panels normalize the ionic current by $\bar{u}\nabla p_0$ to quantify $i$ in relation to the mechanical power input needed to drive it. Inspecting the colormaps indicates that Regime II produces the largest absolute ionic current $i$. However, the optimal operating point depends on the objective. Peak selectivity --- the ability to discriminate between cation and anion fluxes --- is achieved at near-complete EDL overlap ($l_\mathrm{D} \simeq W/2$, i.e.\ $2\tilde{l}_\mathrm{D} \to 1$) and at driving forces just below the flow transition ($\Pi/(2\tilde{l}_\mathrm{D})^2 \lesssim \mathcal{O}(1)$), where electrostatic forces are strong enough to selectively filter ions at the channel constrictions but have not yet been overcome by the mechanical pressure. The domain of high selectivity is broad in the ($\tilde{l}_\mathrm{D}$, $\Pi$) parameter space, making this regime robust to parameter variation. By contrast, the power-normalized ionic current (bottom row of Fig.~\ref{fig:Fig9}) is sharply peaked near the flow transition itself, $\Pi/(2\tilde{l}_\mathrm{D})^2 = \mathcal{O}(1)$, and decays rapidly on either side; energy-efficient current generation therefore requires precise operation near the threshold separating Regime I from Regime II.

To connect these continuum-scale fluxes to individual ion trajectories, we use random walk particle tracking (RWPT) driven by the steady PNPS velocity and electric fields. Each simulation follows $N_\mathrm{p}=10^5$ cations or anions released across a channel constriction in a periodic domain. We sample the velocity distributions from the continuum PNPS model for $c_0=0.005\,\mathrm{M}$ and $\Pi/(2\tilde{l}_\mathrm{D})^2=1.33$ and $7.51$, which lie below and above the flow transition, respectively; the corresponding PNPS solutions are marked by filled symbols in Fig.~\ref{fig:Fig8}. Expressions for the Langevin formulation, particle initialization, and statistical estimators are provided in Appendix~\ref{app:rwpt-methods}.

\begin{figure}[t]
    \centering
    \includegraphics[width=\linewidth]{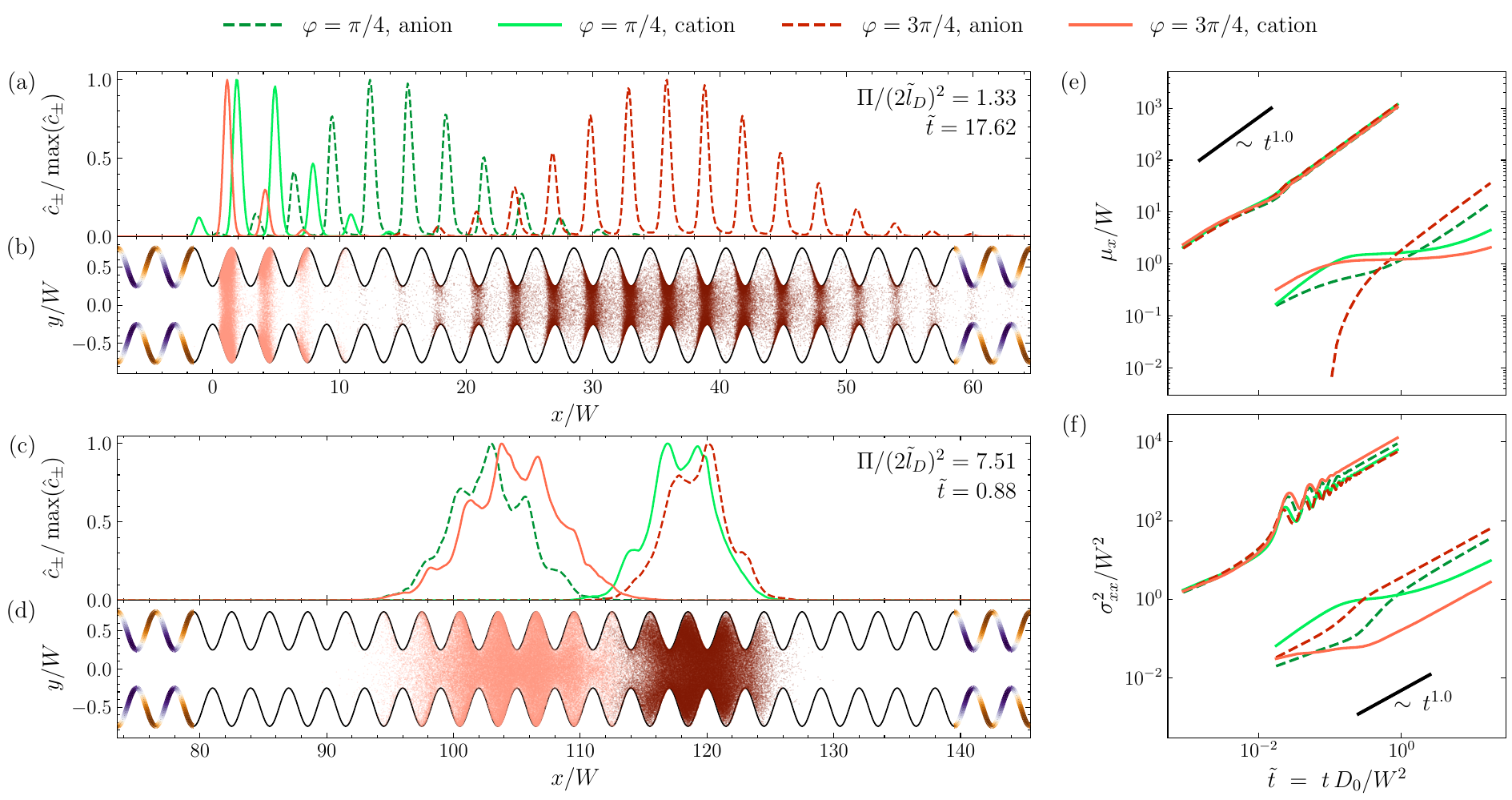}
    \caption[Summary of RWPT ion distributions and longitudinal transport statistics below and above the flow transition.]{RWPT ion distributions and longitudinal transport statistics below (a,b; $\Pi/(2\tilde{l}_\mathrm{D})^2=1.33$ and $\tilde{t}=17.62$) and above (c,d; $\Pi/(2\tilde{l}_\mathrm{D})^2=7.51$ and $\tilde{t}=0.88$) the flow transition for $c_0=0.005\,\mathrm{M}$ ($2\tilde{l}_D=1.64$). The top panels (a,c) show normalized one-dimensional kernel density estimates of the cation (solid) and anion (dashed) distributions for $\pi/4$ and $3\pi/4$. The adjacent, bottom panels (b,d) show representative pore-scale distributions for $\varphi=3\pi/4$; the boundary coloring indicates the sinusoidal surface charge distribution (color convention as in Figs.~\ref{fig:Fig3} and \ref{fig:Fig4}). The right-side panels show the time evolution of (e) the mean displacement and (f) the variance for $\varphi=\pi/4$ (green) and $3\pi/4$ (red); high-pressure-gradient curves are vertically offset by two decades. Solid, light curves (dashed, dark curves) denote statistics for cations (anions).}
    \label{fig:Fig10}
\end{figure}

Figures \ref{fig:Fig10}(a-d) summarize the particle distributions on either side of the flow transition for $\varphi=\pi/4$ and $3\pi/4$ --- \textit{i.e.}, for charge patches shielded from or placed in opposition to the flow entering the pore constriction; the complete $\varphi$-resolved distributions are provided in Figs.~S3 and S4 of the Supplemental Material~\cite{SM}. Near the transition, Fig.~\ref{fig:Fig10}(a) and (b) display the cation and anion density peaks to be centered on successive surface-charge patches. Their patch-to-patch hopping is strongly controlled by $\varphi$. Selectivity is greatest when the charge maximum is placed in opposition to the flow focusing, slightly upstream of the throat ($\varphi=3\pi/4$), whereas both species are better shielded from the strongest advection currents for $\varphi=\pi/4$. Once the flow enters Regime II, Figs.~\ref{fig:Fig10}(c) and (d) show the peaks washing out into broad slugs that access the full channel volume. The dependence on surface-charge placement consequently weakens as mechanical advection overtakes electrokinetic drift.

The right panels, Fig.~\ref{fig:Fig10}(e,f), connect these pore-scale dynamics to effective transport. At late times the longitudinal moments approach $\mu_{\pm,x}\sim\bar{u}_\pm t$ and $\sigma_{\pm,xx}^2\sim2D^\mathrm{eff}_{\pm,xx}t$. Below the transition the mean velocities and dispersion coefficients remain widely separated across ionic species and depend on the choices of $\varphi$, reflecting the selective trapping and release from charge patches. Above the transition, the curves collapse as the ions become mechanically mobilized. Independently calculated RWPT and PNPS selectivities are generally consistent. For example, at $\varphi=3\pi/4$ and $\Pi/(2\tilde{l}_\mathrm{D})^2=1.33$, they give $|\varsigma|=0.95$ and $0.93$, respectively, all RWPT and PNPS selectivities measure at or below $0.1$ at $\Pi/(2\tilde{l}_\mathrm{D})^2=7.51$. The main exception is $\varphi=\pi/4$ near the transition, where the cation and anion flux curves cross and small differences between the methods strongly affect $\varsigma$. The complete set of transport statistics and comparison between continuum and discrete modeling outputs are reported in Table~\ref{tab:RWPT_statistics}.

We note that the parameters of our model were not optimized for selectivity. Further tuning of $c_0$, $L$, $\delta W$, $b$, $\sigma_0$, and $\varphi$, together with adjustments to the geometric and surface-charge waveforms, will likely produce larger currents with enhanced selectivity and rectification.

\section{Conclusions and perspectives}
\label{sec:conclusion}

Actively or chemically controllable pores have emerged as a promising strategy for regulating transport in nanofluidic systems \cite{tsutsui2026chemistry, tagliazucchi2011ion}. Understanding fluid flow and scalar mixing in confined geometries is therefore of fundamental importance across physics, biology, hydrology, and engineering, as it underpins the upscaling of microscale transport mechanisms to macroscopic behavior. In this work, we presented a numerical investigation of electrokinetic flow and ion transport in charge-patterned, corrugated nanochannels by solving the fully coupled Poisson-Nernst-Planck-Stokes (PNPS) equations under both pressure-driven and electrically-driven forcing conditions.

Our results demonstrate that the interplay between surface charge patterning, geometric corrugation, and electrolyte composition gives rise to distinct flow regimes governed by the competition between electrostatic and mechanical forces. Under an applied pressure gradient, the system exhibits a transition from an electrokinetically inhibited regime --- in which electrostatic forces resist ion displacement from the electric double layer and suppress throughput --- to a mechanically dominated regime characterized by Poiseuille-like flow. This transition is controlled by the relative magnitude of electrostatic and pressure forces and may occur abruptly, leading to a nonlinear, threshold-like increase in flowrate. Under an applied electric field, a qualitatively different behavior emerges: transitioning between low and high field strengths --- where the local electrostatic force due to the streaming potential acting on the ions is either strong or weak relative to the driving field --- causes a reversal in the direction of net flow. At high electric field strengths, the flowrate exhibits a nonlinear scaling with field amplitude, and this nonlinearity is most pronounced at low salt concentrations, where electrostatic effects are least screened.

A central result is that phase shifts between the surface charge distribution and the channel geometry provide a robust mechanism to control both the magnitude and direction of ionic fluxes. In particular, we demonstrate the emergence of rectified, diode-like transport behavior, whereby finite pressure gradients asymmetrically activate low- and high-flow states depending on the charge distribution. This mechanism arises from the preferential release and trapping of counterions within the electric double layer and is enhanced by geometric confinement.

In contrast to previous studies that considered uniform or weakly perturbed charge distributions or relied on linearized electrokinetic models, the present work captures the fully nonlinear coupling between ion transport, electrostatics, and hydrodynamics in corrugated geometries. While \citet{curk2024discontinuous} identified gating transitions in periodically charged channels, our results show that the combined effects of geometric corrugation and phase-shifted charge patterning introduce an additional symmetry-breaking mechanism that enables directional flow generation and selective ion transport under both pressure-driven and electrically-driven conditions.

All simulations presented here are two-dimensional, representing flow in a slit-like channel geometry. Real nanofluidic devices --- including track-etched nanopores, anodized alumina channels, and carbon nanotubes --- confine the electrolyte in two transverse dimensions simultaneously, which modifies the cross-sectional velocity profile and introduces EDL overlap along two walls rather than one~\cite{bocquet2010nanofluidics,secchi2016massive}. Corner effects and the distribution of surface charge over a cylindrical or rectangular perimeter will quantitatively alter flowrates and ionic currents for a given applied driving force. Nevertheless, the two-regime structure --- governed by the ratio of pressure to electrostatic drift force, $\Pi/(2\tilde{l}_\mathrm{D})^2$ --- reflects a competition between bulk mechanical and interfacial electrostatic forces that is independent of cross-sectional shape, and the phase-controlled rectification mechanism relies on axial symmetry breaking between surface charge and channel geometry, a feature preserved in three-dimensional channels with axially varying cross-sections. Extending the present framework to rectangular or cylindrical geometries represents a natural and tractable next step.

These findings provide a physical framework for designing nanofluidic systems with tunable transport properties, with potential applications in ion-selective membranes, energy harvesting, and subsurface transport processes such as carbon sequestration and brine mineral recovery. The present continuum treatment, however, neglects finite ion size effects and inter-ionic correlations, which become increasingly important as confinement approaches molecular length scales. Additionally, all simulations assume equal diffusion coefficients for cations and anions, $D_+ = D_- = D_0$. For a NaCl electrolyte, the diffusivity of Na$^+$ ($\approx 1.33 \times 10^{-9}$ m$^2$\,s$^{-1}$) is roughly 35\% lower than that of Cl$^-$ ($\approx 2.03 \times 10^{-9}$ m$^2$\,s$^{-1}$); this asymmetry generates a diffusion potential that modifies the local electric field and can quantitatively alter ionic current and charge selectivity. The qualitative flow regimes and transition behavior identified here are expected to persist for unequal diffusivities, but quantitative predictions of selectivity and rectification --- particularly near the Regime I--Regime II transition --- may be affected. Incorporating steric exclusion and ion-ion correlations, through modified PNP formulations~\cite{borukhov1997steric}, classical density functional theory~\cite{petersen2024toward}, or molecular dynamics simulations, represents an important avenue for future work. For charge-regulating surfaces, an additional refinement to the model would replace the prescribed surface-charge distribution with a self-consistent surface-chemistry relation, allowing the local charge and slip length to respond to electrolyte composition, electrostatic potential, and wall curvature, while also accounting for surface conduction along the interface.

A compelling extension is the design of dynamically fluctuating or wrinkling geometries that mechanically gate flow as a function of velocity, extending the perturbative framework of \citet{marbach2018transport, marbach2019active} to charged nanochannels. 

\begin{acknowledgments}
Thomas Petersen thanks Landon Allen, whose steadfast collaboration through USC's Center for Undergraduate Research in Viterbi Engineering both inspired and progressed the study. This work was supported with start-up funds from the Viterbi School of Engineering at the University of Southern California. Felipe P. J. de Barros acknowledges the partial support from the NSF (Award Number 2333378).
\end{acknowledgments}

\appendix

\section{Linearized response in a flat channel with sinusoidal surface charge}
\label{app:LRT-solution}

Using the stagnant equilibrium solution of the linearized Poisson--Boltzmann equation~(\ref{eq:equilibrium_potential}) as a reference state, \citet{ajdari1995electro,ajdari1996generation} performed a perturbation expansion of the PNPS equations --- choosing $\delta \bm{E}=\bm{E}_\mathrm{ext}$ as the perturbation --- to solve for electric field-driven flow. Here, we adapt Ajdari's solution to the case of slip-flow boundary conditions.

Assuming $\bm{E}\approx -\nabla\phi^\mathrm{eq}+\bm{E}_\mathrm{ext}$ and neglecting any streaming of the ion clouds, $\rho_\mathrm{c}\approx\rho_\mathrm{c}^\mathrm{eq}$, the Stokes equation~(\ref{eq:momentum_balance}) and incompressibility provide
\begin{equation}\label{eq:streamfunction}
 \mu \nabla^4\psi =\nabla \rho_\mathrm{c}^\mathrm{eq}\times E_\mathrm{ext}\mathbf{e}_x = \varepsilon E_\mathrm{ext}(\partial_{yxx}\phi^\mathrm{eq} +\partial_{yyy}\phi^\mathrm{eq}),
\end{equation}
where $\psi$ is the streamfunction that grants access to the velocity components through $\partial_y\psi=u$ and $\partial_x\psi=-v$. The particular solution of Eq.~(\ref{eq:streamfunction}) is
\begin{equation}
\psi_\mathrm{p}(x,y) = \frac{\sigma_0 E_\mathrm{ext}l_\mathrm{D}}{\mu}\sin(qx)\frac{\sinh(Ky)}{\sinh(Kw)},
\end{equation}
using $w=W/2$ to simplify notation, while the homogeneous solution is sought in the form
\begin{equation}
\begin{split}
\psi_\mathrm{h}(x,y)&=\frac{\sigma_0 E_\mathrm{ext}l_\mathrm{D}}{\mu}\sin(qx)\\
&\times\left[A_q\cosh(qy) + B_q\sinh(qy) + C_q\, y\cosh(qy) + D_q\, y\sinh(qy)\right].
\end{split}
\end{equation}
The constants $A_q$, $B_q$, $C_q$, and $D_q$ are calculated using the velocity boundary conditions along $y=0$ and $y=w$. Symmetry requires $v(y=0)=-\partial_x\psi|_{y=0}=0$ and $\partial_yu|_{y=0}=\partial_{yy}\psi|_{y=0}=0$, which sets $A_q=D_q=0$. The no-flow condition $v(y=W/2)=-\partial_x\psi|_{y=W/2}=0$ and slip-flow condition $\partial_yu|_{y=W/2}+bu(y=W/2)=-(\partial_{yy}\psi+b\partial_y\psi)|_{y=W/2}=0$ determine $B_q$ and $C_q$:
\begin{subequations}\label{eq:LR-coefficients}
\begin{align}
B_q &= \frac{2\left(\cosh(wq)\left[1 - bw(K^2 - q^2) - wK\coth(wK)\right] + (2b + w)q\sinh(wq)\right)}{2(b+w)q - 2bq\cosh(2wq) - \sinh(2wq)}
\label{eq:LR-coefficient-B}\\
C_q &= \frac{q - \left[b(K^2 - q^2) + K\coth(w K)\right]\tanh(w q)}{\cosh(w q)\left[\tanh(wq) + (2b + w)q\tanh^2(wq) - wq\right]}.
\label{eq:LR-coefficient-C}
\end{align}
\end{subequations}
The dimensionless function appearing in Eq.~(\ref{eq:LR-velocities}) is
\begin{equation}\label{eq:LR-g}
g(y) = \frac{\sinh(Ky)}{\sinh(Kw)} + B_q\sinh(qy) + C_q y \cosh(q y).
\end{equation}
Substitution into Eqs.~(\ref{eq:LR-velocities-u}) and (\ref{eq:LR-velocities-v}) yields the velocity components used in the main-text.

\section{Random-walk particle tracking method and transport statistics}
\label{app:rwpt-methods}

\subsection{RWPT formulation and simulation setup}

Using the steady-state velocity profile, $\bm{v}$, and electric field, $\bm{E}$, produced by the numerical solution of Eqs.~(\ref{eq:governing_equations}), we simulate the trajectories of point charges, $\bm{r}_\pm(t|\bm{r}_{\pm,0})$, with the Langevin equation~\cite{gardiner2009stochastic}
\begin{equation}\label{eq:Langevin}
   \frac{\mathrm{d} \bm{r}_\pm(t|\bm{r}_{\pm,0})}{\mathrm{d} t}= \bm{v}[\bm{r}_\pm(t|\bm{r}_{\pm,0})] + M_\pm z_\pm \bm{E}[\bm{r}_\pm(t|\bm{r}_{\pm,0})] + \bm{\xi}(t).
\end{equation}
The thermal fluctuations are represented by two-dimensional Gaussian white noise, $\bm{\xi}(t)$, with $\langle \xi_j(t)\rangle=0$ and $\langle \xi_i(t)\xi_j(\tau)\rangle=2D_0\delta_{ij}\delta(t-\tau)$.

Each simulation begins with $N_\mathrm{p}=10^5$ cations or anions uniformly distributed along a line source of length $\ell_\mathrm{p}\ll L$ spanning the channel constriction at $x=0$. Their trajectories are advanced in parallel by solving Eq.~(\ref{eq:Langevin}) with the GPU-accelerated, open-source \texttt{PAR$^2$} simulator~\cite{rizzo2019par2}. The domain is periodic, and particle positions are unwrapped to record the number of channel wavelengths traversed. We sample the PNPS fields for $k=1$, $c_0=0.005\,\mathrm{M}$ ($2\tilde l_\mathrm{D}=1.64$), and $\Pi/(2\tilde l_\mathrm{D})^2=1.33$ or $7.51$, corresponding to states below and above the flow transition. These PNPS solutions are marked by filled symbols in the left panels of Fig.~\ref{fig:Fig8}.

\subsection{Distribution and effective-transport statistics}

We calculate one-dimensional estimates of the ion concentration fields using the kernel density estimate
\begin{equation}\label{eq:KDE}
  \hat{c}_\pm(\tilde{x})
  = \frac{1}{\mathcal{A}(\tilde{x})\tilde{h}\sqrt{2\pi}}
  \sum_{i=1}^{N_\mathrm{p}}
  \exp\left[-\frac{(\tilde{x}-\tilde{r}_{\pm,x}^{i})^2}{2\tilde{h}^2}\right],
\end{equation}
where $\{\tilde{r}^i_{\pm,x}\}$ are the dimensionless longitudinal particle positions, $\mathcal{A}(\tilde{x})$ is the local channel width, and $\tilde{h}=0.05L/W$ is the smoothing bandwidth. This bandwidth is small relative to the charge-patch spacing, $\tilde{L}/(2k)$, but sufficiently large to provide smooth profiles for the finite particle ensemble. Equation~(\ref{eq:KDE}) is evaluated on a uniform grid and normalized by $\max[\hat c_\pm(\tilde x)]$ for visualization.

After the initially concentrated plume assumes its transverse steady-state distribution, the long-time rates of change of its first two longitudinal moments define the mean particle velocity and effective dispersion coefficient:
\begin{subequations}\label{eq:RWPT_moments}
\begin{align}
\bar{u}_\pm
&=\lim_{t\to\infty}\frac{\mathrm{d}}{\mathrm{d}t}\langle r_{\pm,x}\rangle
=\lim_{t\to\infty}\frac{\mathrm{d}\mu_{\pm,x}}{\mathrm{d}t},\\
D^\mathrm{eff}_{\pm,xx}
&=\lim_{t\to\infty}\frac{1}{2}\frac{\mathrm{d}}{\mathrm{d}t}
\left(\langle r_{\pm,x}^{2}\rangle-\langle r_{\pm,x}\rangle^2\right)
=\lim_{t\to\infty}\frac{1}{2}\frac{\mathrm{d}\sigma_{\pm,xx}^{2}}{\mathrm{d}t}.
\end{align}
\end{subequations}
Here $r_{\pm,x}=\bm{r}_\pm\cdot\mathbf{e}_x$, while $\mu_{\pm,x}$ and $\sigma_{\pm,xx}^2$ denote the mean and variance of the unwrapped particle positions. Table~\ref{tab:RWPT_statistics} reports these effective statistics for all simulated phase offsets. Because the $k=1$ configuration is net-neutral, the RWPT selectivity follows directly from the mean velocities, $\varsigma=(\bar u_+-\bar u_-)/(\bar u_++\bar u_-)$.

\begin{table}[t]
\caption{Ion transport statistics from RWPT simulations. The $|\varsigma|$ values in parentheses are the corresponding measurements from the PNPS output.}
\label{tab:RWPT_statistics}
\begin{ruledtabular}
\begin{tabular}{ccccccc}
$\Pi/(2 \tilde{l}_\mathrm{D})^2$ & $\varphi$ &
$\bar{u}_+ W/D_0$ &
$\bar{u}_- W/D_0$ &
$D_{+,xx}^\mathrm{eff}/D_0$ & $D_{-,xx}^\mathrm{eff}/D_0$ & $|\varsigma|$ \\
\hline
$1.33$ & $0$       & $1.94\times 10^{0}$ & $3.43 \times 10^{-1}$ & $1.82\times 10^{0}$ & $4.75\times 10^{-1}$ & 0.70 (0.79) \\
$1.33$ & $\pi/4$   & $1.72 \times 10^{-1}$ & $8.34 \times 10^{-1}$ & $2.51\times 10^{-1}$ & $1.02\times 10^{0}$ & 0.66 (0.30) \\
$1.33$ & $\pi/2$   & $2.54 \times 10^{-2}$ & $1.34 \times 10^{0}$ & $3.78\times 10^{-2}$ & $1.33\times 10^{0}$ & 0.96 (0.90) \\
$1.33$ & $3\pi/4$  & $5.18 \times 10^{-2}$ & $2.07 \times 10^{0}$ & $7.69\times 10^{-2}$ & $1.78 \times 10^{0}$ & 0.95 (0.93) \\
$7.51$ & $0$       & $1.39 \times 10^{2}$ & $1.16 \times 10^{2}$ & $3.03 \times 10^{0}$ & $6.84 \times 10^{0}$ & 0.09 (0.10) \\
$7.51$ & $\pi/4$   & $1.34 \times 10^{2}$ & $1.17 \times 10^{2}$ & $3.64 \times 10^{0}$ & $5.21 \times 10^{0}$ & 0.07 (0.06) \\
$7.51$ & $\pi/2$   & $1.26 \times 10^{2}$ & $1.25 \times 10^{2}$ & $5.51 \times 10^{0}$ & $3.96 \times 10^{0}$ & 0.00 (0.02) \\
$7.51$ & $3\pi/4$  & $1.19 \times 10^{2}$ & $1.35 \times 10^{2}$ & $7.39 \times 10^{0}$ & $3.22 \times 10^{0}$ & 0.06 (0.09) \\
\end{tabular}
\end{ruledtabular}
\end{table}

The table highlights the orders-of-magnitude increase in mean ion velocity and the accompanying loss of selectivity across the transition. Below the transition, counterions protected in the channel troughs can move and spread more slowly than Brownian motion alone, with normalized velocities or dispersion coefficients below unity. Above the transition, mechanical advection mobilizes both species and reduces $|\varsigma|$ below $0.1$. RWPT and PNPS selectivities agree closely except for $\varphi=\pi/4$ at $\Pi/(2\tilde l_\mathrm{D})^2=1.33$. At this point the cation and anion flux curves cross [Fig.~\ref{fig:Fig8}(d)], making the selectivity particularly sensitive to small differences between the two methods.

\bibliography{references}

@article{thompson1982boundary,
  title={Boundary-fitted coordinate systems for numerical solution of partial differential equations---{A} review},
  author={Thompson, Joe F. and Warsi, Zahir U. A. and Mastin, C. Wayne},
  journal={Journal of Computational Physics},
  volume={47},
  number={1},
  pages={1--108},
  year={1982},
  publisher={Elsevier},
  doi={10.1016/0021-9991(82)90066-3}
}

@article{vanderheyden2006electrokinetic,
  title   = {Electrokinetic Energy Conversion Efficiency in Nanofluidic Channels},
  author  = {van der Heyden, Frank H. J. and Bonthuis, Douwe Jan and Stein, Derek and Meyer, Christine and Dekker, Cees},
  journal = {Nano Letters},
  volume  = {6},
  number  = {10},
  pages   = {2232--2237},
  year    = {2006},
  doi     = {10.1021/nl061524l}
}

@article{vanderheyden2007power,
  title   = {Power Generation by Pressure-Driven Transport of Ions in Nanofluidic Channels},
  author  = {van der Heyden, Frank H. J. and Bonthuis, Douwe Jan and Stein, Derek and Meyer, Christine and Dekker, Cees},
  journal = {Nano Letters},
  volume  = {7},
  number  = {4},
  pages   = {1022--1025},
  year    = {2007},
  doi     = {10.1021/nl070194h}
}

@article{behrens2001charge,
  title   = {The Charge of Glass and Silica Surfaces},
  author  = {Behrens, Sven H. and Grier, David G.},
  journal = {The Journal of Chemical Physics},
  volume  = {115},
  number  = {14},
  pages   = {6716--6721},
  year    = {2001},
  doi     = {10.1063/1.1404988}
}

@article{doyle1998structure,
  title={The structure of the potassium channel: Molecular basis of {K}$^+$ conduction and selectivity},
  author={Doyle, Declan A and Morais Cabral, Jo{\~a}o and Pfuetzner, Richard A and Kuo, Anling and Gulbis, Jacqueline M and Cohen, Steven L and Chait, Brian T and MacKinnon, Roderick},
  journal={Science},
  volume={280},
  number={5360},
  pages={69--77},
  year={1998},
  publisher={American Association for the Advancement of Science},
  doi={10.1126/science.280.5360.69}
}

@article{borukhov1997steric,
  title={Steric effects in electrolytes: A modified {P}oisson--{B}oltzmann equation},
  author={Borukhov, Itamar and Andelman, David and Orland, Henri},
  journal={Physical Review Letters},
  volume={79},
  number={3},
  pages={435},
  year={1997},
  publisher={APS},
  doi={10.1103/PhysRevLett.79.435}
}

@article{ghosal2002lubrication,
  title={Lubrication theory for electro-osmotic flow in a microfluidic channel of slowly varying cross-section and wall charge},
  author={Ghosal, Sandip},
  journal={Journal of Fluid Mechanics},
  volume={459},
  pages={103--128},
  year={2002},
  publisher={Cambridge University Press},
  doi={10.1017/S0022112002007899}
}

@article{schoch2005effect,
  pages={100604},
  title={Effect of the surface charge on ion transport through nanoslits},
  author={Schoch, Reto B. and van Lintel, Harald and Renaud, Philippe},
  journal={Physics of Fluids},
  volume={17},
  number={10},
  year={2005},
  publisher={AIP Publishing},
  doi={10.1063/1.1896936}
}

@article{secchi2016massive,
  title={Massive radius-dependent flow slippage in carbon nanotubes},
  author={Secchi, Eleonora and Marbach, Sophie and Nigu{\`e}s, Antoine and Stein, Derek and Siria, Alessandro and Bocquet, Lyd{\'e}ric},
  journal={Nature},
  volume={537},
  number={7619},
  pages={210--213},
  year={2016},
  publisher={Nature Publishing Group UK London},
  doi={10.1038/nature19315}
}

@article{davis1978surface,
  title={Surface ionization and complexation at the oxide/water interface: {I}. {C}omputation of electrical double layer properties in simple electrolytes},
  author  = {Davis, James A. and James, Robert O. and Leckie, James O.},
  journal = {Journal of Colloid and Interface Science},
  volume  = {63},
  number  = {3},
  pages   = {480--499},
  year    = {1978},
  doi     = {10.1016/S0021-9797(78)80009-5}
}

@article{trefalt2016charge,
  title   = {Charge Regulation in the Electrical Double Layer:
             Ion Adsorption and Surface Interactions},
  author  = {Trefalt, Gregor and Behrens, Sven Holger and
             Borkovec, Michal},
  journal = {Langmuir},
  volume  = {32},
  number  = {2},
  pages   = {380--400},
  year    = {2016},
  doi     = {10.1021/acs.langmuir.5b03611}
}

@article{vlassiouk2007nanofluidic,
  doi={10.1021/nl062924b},
  title={Nanofluidic diode},
  author={Vlassiouk, Ivan and Siwy, Zuzanna S},
  journal={Nano Letters},
  volume={7},
  number={3},
  pages={552--556},
  year={2007},
  publisher={ACS Publications}
}

@article{messinger2010suppression,
  title   = {Suppression of Electro-Osmotic Flow by Surface Roughness},
  author  = {Messinger, Robert J. and Squires, Todd M.},
  journal = {Physical Review Letters},
  volume  = {105},
  number  = {14},
  pages   = {144503},
  year    = {2010},
  doi     = {10.1103/PhysRevLett.105.144503}
}

@article{wang2010electrochemical,
  title   = {Electrochemical Charge of Silica Surfaces at High Ionic
             Strength in Narrow Channels},
  author  = {Wang, Moran and Revil, Andr{\'e}},
  journal = {Journal of Colloid and Interface Science},
  volume  = {343},
  number  = {1},
  pages   = {381--386},
  year    = {2010},
  doi     = {10.1016/j.jcis.2009.11.039}
}

@article{tagliazucchi2011ion,
  title={Ion transport and molecular organization are coupled in polyelectrolyte-modified nanopores},
  author={Tagliazucchi, Mario and Rabin, Yitzhak and Szleifer, Igal},
  journal={Journal of the American Chemical Society},
  volume={133},
  number={44},
  pages={17753--17763},
  year={2011},
  publisher={ACS Publications},
  doi={10.1021/ja2063605}
}

@article{tsutsui2026chemistry,
  title={Chemistry-driven autonomous nanopore membranes},
  author={Tsutsui, Makusu and Hsu, Wei-Lun and Garoli, Denis and Douaki, Ali and Komoto, Yuki and Daiguji, Hirofumi and Kawai, Tomoji},
  journal={Nature Communications},
  volume={17},
  number={1},
  pages={1496},
  year={2026},
  publisher={Nature Publishing Group UK London},
  doi={10.1038/s41467-026-68800-x}
}

@article{harlow1965numerical,
  title={Numerical calculation of time-dependent viscous incompressible flow of fluid with free surface},
  author={Harlow, Francis H. and Welch, J. Eddie},
  journal={Physics of Fluids},
  volume={8},
  number={12},
  pages={2182--2189},
  year={1965},
  doi={10.1063/1.1761178}
}

@article{goyal2024generalizing,
  title={Generalizing electroosmotic-flow predictions over charge-modulated periodic topographies: tuneable far-field effects},
  author={Goyal, Vishal and Datta, Subhra and Chakraborty, Suman},
  journal={Journal of Fluid Mechanics},
  volume={990},
  pages={A1},
  year={2024},
  publisher={Cambridge University Press},
  doi={10.1017/jfm.2024.491}
}

@article{khair2008fundamental,
  pages={087102},
  title={Fundamental aspects of concentration polarization arising from nonuniform electrokinetic transport},
  author={Khair, Aditya S and Squires, Todd M},
  journal={Physics of Fluids},
  volume={20},
  number={8},
  year={2008},
  publisher={AIP Publishing},
  doi={10.1063/1.2963507}
}

@book{probstein2005physicochemical,
  title={Physicochemical Hydrodynamics: An Introduction},
  author={Probstein, Ronald F.},
  edition={2nd},
  year={1994},
  publisher={John Wiley \& Sons},
  address={New York}
}

@article{bocquet2007flow,
  title={Flow boundary conditions from nano- to micro-scales},
  author={Bocquet, Lyd{\'e}ric and Barrat, Jean-Louis},
  journal={Soft Matter},
  volume={3},
  number={6},
  pages={685--693},
  year={2007},
  publisher={Royal Society of Chemistry},
  doi={10.1039/b616490k}
}

@article{bocquet2010nanofluidics,
  title={Nanofluidics, from bulk to interfaces},
  author={Bocquet, Lyd{\'e}ric and Charlaix, Elisabeth},
  journal={Chemical Society Reviews},
  volume={39},
  number={3},
  pages={1073--1095},
  year={2010},
  publisher={Royal Society of Chemistry},
  doi={10.1039/b909366b}
}

@article{lanyon2007recessed,
  title={Recessed nanoband electrodes fabricated by focused ion beam milling},
  author={Lanyon, Yvonne H. and Arrigan, Damien W. M.},
  journal={Sensors and Actuators B: Chemical},
  volume={121},
  number={1},
  pages={341--347},
  year={2007},
  publisher={Elsevier},
  doi={10.1016/j.snb.2006.11.029}
}

@article{li2001ion,
  title={Ion-beam sculpting at nanometre length scales},
  author={Li, Jiali and Stein, Derek and McMullan, Ciaran and Branton, Daniel and Aziz, Michael J and Golovchenko, Jene A},
  journal={Nature},
  volume={412},
  number={6843},
  pages={166--169},
  year={2001},
  publisher={Nature Publishing Group UK London},
  doi={10.1038/35084037}
}

@article{carusela2025transport,
  title={Transport of electrolytes across nanochannels: The role of slip},
  author={Carusela, M Florencia and Harting, Jens and Malgaretti, Paolo},
  journal={Physics of Fluids},
  volume={37},
  number={5},
  pages={052005},
  year={2025},
  publisher={AIP Publishing},
  doi={10.1063/5.0256681}
}

@article{malgaretti2019driving,
  pages={084902},
  title={Driving an electrolyte through a corrugated nanopore},
  author={Malgaretti, Paolo and Janssen, Mathijs and Pagonabarraga, Ignacio and Rub{\'\i}, J. Miguel},
  journal={The Journal of Chemical Physics},
  volume={151},
  number={8},
  year={2019},
  publisher={AIP Publishing},
  doi={10.1063/1.5110349}
}

@article{park2006eddies,
  title={Eddies in a bottleneck: An arbitrary {D}ebye length theory for capillary electroosmosis},
  author={Park, Stella Y and Russo, Christopher J and Branton, Daniel and Stone, Howard A},
  journal={Journal of Colloid and Interface Science},
  volume={297},
  number={2},
  pages={832--839},
  year={2006},
  publisher={Elsevier},
  doi={10.1016/j.jcis.2005.11.045}
}

@article{Yoon_Dentz_Kang_2021, 
title={Optimal fluid stretching for mixing-limited reactions in rough channel flows}, volume={916}, DOI={10.1017/jfm.2021.208}, 
journal={Journal of Fluid Mechanics}, 
author={Yoon, Seonkyoo and Dentz, Marco and Kang, Peter K.}, 
year={2021}, 
pages={A45}}

@book{gardiner2009stochastic,
  address={Berlin},
  edition={4th},
  title={Stochastic Methods: A Handbook for the Natural and Social Sciences},
  author={Gardiner, Crispin W.},
  year={2009},
  publisher={Springer}
}

@article{basilio2022linking,
  title={Linking mixing and flow topology in porous media: An experimental proof},
  author={Basilio Hazas, M. and Ziliotto, F. and Rolle, M. and Chiogna, G.},
  journal={Physical Review E},
  volume={105},
  number={3},
  pages={035105},
  year={2022},
  publisher={APS},
  doi={10.1103/PhysRevE.105.035105}
}

@article{sapp2024deionization,
  title={Deionization shock waves and ionic separations in heterogeneous porous media},
  author={Sapp, Alexander D and Tian, Huanhuan and Bazant, Martin Z},
  journal={Physical Review Fluids},
  volume={9},
  number={7},
  pages={073701},
  year={2024},
  publisher={APS},
  doi={10.1103/PhysRevFluids.9.073701}
}

@article{burgreen1964electrokinetic,
  title={Electrokinetic flow in ultrafine capillary slits},
  author={Burgreen, D. and Nakache, F. R.},
  journal={The Journal of Physical Chemistry},
  volume={68},
  number={5},
  pages={1084--1091},
  year={1964},
  publisher={ACS Publications},
  doi={10.1021/j100787a019}
}

@article{webb1971heat,
  title={Heat transfer and friction in tubes with repeated-rib roughness},
  author={Webb, R. L. and Eckert, E. R. G. and Goldstein, R. J.},
  journal={International Journal of Heat and Mass Transfer},
  volume={14},
  number={4},
  pages={601--617},
  year={1971},
  publisher={Elsevier},
  doi={10.1016/0017-9310(71)90009-3}
}

@article{acar1993principles,
  title={Principles of electrokinetic remediation},
  author={Acar, Yalcin B and Alshawabkeh, Akram N},
  journal={Environmental Science \& Technology},
  volume={27},
  number={13},
  pages={2638--2647},
  year={1993},
  publisher={ACS Publications},
  doi={10.1021/es00049a002}
}

@article{probstein1993removal,
  title={Removal of contaminants from soils by electric fields},
  author={Probstein, Ronald F and Hicks, R Edwin},
  journal={Science},
  volume={260},
  number={5107},
  pages={498--503},
  year={1993},
  publisher={American Association for the Advancement of Science},
  doi={10.1126/science.260.5107.498}
}

@article{thomas2008enhanced,
  title={Enhanced oil recovery---{A}n overview},
  author={Thomas, Sara},
  journal={Oil \& Gas Science and Technology -- Revue de l'IFP},
  volume={63},
  number={1},
  pages={9--19},
  year={2008},
  publisher={EDP Sciences},
  doi={10.2516/ogst:2007060}
}

@article{epsztein2020towards,
  title={Towards single-species selectivity of membranes with subnanometre pores},
  author={Epsztein, Razi and DuChanois, Ryan M and Ritt, Cody L and Noy, Aleksandr and Elimelech, Menachem},
  journal={Nature Nanotechnology},
  volume={15},
  number={6},
  pages={426--436},
  year={2020},
  publisher={Nature Publishing Group UK London},
  doi={10.1038/s41565-020-0713-6}
}

@article{deng2015water,
  title={Water purification by shock electrodialysis: Deionization, filtration, separation, and disinfection},
  author={Deng, Daosheng and Aouad, Wassim and Braff, William A and Schlumpberger, Sven and Suss, Matthew E and Bazant, Martin Z},
  journal={Desalination},
  volume={357},
  pages={77--83},
  year={2015},
  publisher={Elsevier},
  doi={10.1016/j.desal.2014.11.011}
}

@article{ajdari2000pumping,
  title={Pumping liquids using asymmetric electrode arrays},
  author={Ajdari, Armand},
  journal={Physical Review E},
  volume={61},
  number={1},
  pages={R45},
  year={2000},
  publisher={APS},
  doi={10.1103/PhysRevE.61.R45}
}

@article{curk2024discontinuous,
  title={Discontinuous transition in electrolyte flow through charge-patterned nanochannels},
  author={Curk, Tine and Leyva, Sergi G and Pagonabarraga, Ignacio},
  journal={Physical Review Letters},
  volume={133},
  number={7},
  pages={078201},
  year={2024},
  publisher={APS},
  doi={10.1103/PhysRevLett.133.078201}
}

@article{ramos1998ac,
  title={{AC} electrokinetics: A review of forces in microelectrode structures},
  author={Ramos, Antonio and Morgan, Hywel and Green, Nicolas G and Castellanos, Antonio},
  journal={Journal of Physics D: Applied Physics},
  volume={31},
  number={18},
  pages={2338},
  year={1998},
  publisher={IOP Publishing},
  doi={10.1088/0022-3727/31/18/021}
}

@article{rizzo2019par2,
  title={{PAR}$^2$: Parallel random walk particle tracking method for solute transport in porous media},
  author={Rizzo, C. B. and Nakano, A. and de Barros, F. P. J.},
  journal={Computer Physics Communications},
  volume={239},
  pages={265--271},
  year={2019},
  publisher={Elsevier},
  doi={10.1016/j.cpc.2019.01.013}
}

@article{mohammadi2013pressure,
  title={Pressure losses in grooved channels},
  author={Mohammadi, A and Floryan, Jerzy M},
  journal={Journal of Fluid Mechanics},
  volume={725},
  pages={23--54},
  year={2013},
  publisher={Cambridge University Press},
  doi={10.1017/jfm.2013.184}
}

@article{rolle2018nernst,
  title={{N}ernst--{P}lanck-based description of transport, {C}oulombic interactions, and geochemical reactions in porous media: Modeling approach and benchmark experiments},
  author={Rolle, M. and Sprocati, R. and Masi, M. and Jin, B. and Muniruzzaman, M.},
  journal={Water Resources Research},
  volume={54},
  number={4},
  pages={3176--3195},
  year={2018},
  publisher={Wiley Online Library},
  doi={10.1002/2017WR022344}
}

@article{sprocati2022,
  title={On the interplay between electromigration and electroosmosis during electrokinetic transport in heterogeneous porous media},
  author={Sprocati, R. and Rolle, M.},
  journal={Water Research},
  volume={213},
  pages={118161},
  year={2022},
  publisher={Elsevier},
  doi={10.1016/j.watres.2022.118161}
}

@article{henri2015,
  title={Probabilistic human health risk assessment of degradation-related chemical mixtures in heterogeneous aquifers: Risk statistics, hot spots, and preferential channels},
  author={Henri, C. V. and Fern{\`a}ndez-Garcia, D. and de Barros, F. P. J.},
  journal={Water Resources Research},
  volume={51},
  number={6},
  pages={4086--4108},
  year={2015},
  publisher={Wiley Online Library},
  doi={10.1002/2014WR016717}
}

@article{dentz2023review,
  title={Mixing in porous media: concepts and approaches across scales},
  author={Dentz, M. and Hidalgo, J. J. and Lester, D.},
  journal={Transport in Porous Media},
  volume={146},
  number={1},
  pages={5--53},
  year={2023},
  publisher={Springer},
  doi={10.1007/s11242-022-01852-x}
}

@article{squires2005microfluidics,
  title={Microfluidics: Fluid physics at the nanoliter scale},
  author={Squires, Todd M and Quake, Stephen R},
  journal={Reviews of Modern Physics},
  volume={77},
  number={3},
  pages={977--1026},
  year={2005},
  publisher={APS},
  doi={10.1103/RevModPhys.77.977}
}

@article{reguera2001kinetic,
  title={Kinetic equations for diffusion in the presence of entropic barriers},
  author={Reguera, David and Rub{\'\i}, J. Miguel},
  journal={Physical Review E},
  volume={64},
  number={6},
  pages={061106},
  year={2001},
  publisher={APS},
  doi={10.1103/PhysRevE.64.061106}
}

@article{ling2024,
  title={Dispersion control in coupled channel-heterogeneous porous media systems},
  author={Ling, B. and Shan, R. and de Barros, F. P. J.},
  journal={Physical Review Fluids},
  volume={9},
  number={6},
  pages={064502},
  year={2024},
  publisher={APS},
  doi={10.1103/PhysRevFluids.9.064502}
}

@article{shipley2010,
  title={Multiscale modelling of fluid and drug transport in vascular tumours},
  author={Shipley, R. J. and Chapman, S. J.},
  journal={Bulletin of Mathematical Biology},
  volume={72},
  pages={1464--1491},
  year={2010},
  publisher={Springer},
  doi={10.1007/s11538-010-9504-9}
}

@article{sanaei2017,
  title={Flow and fouling in membrane filters: effects of membrane morphology},
  author={Sanaei, P. and Cummings, L. J.},
  journal={Journal of Fluid Mechanics},
  volume={818},
  pages={744--771},
  year={2017},
  publisher={Cambridge University Press},
  doi={10.1017/jfm.2017.102}
}

@article{ling2018,
  pages={076601},
  title={Hydrodynamic dispersion in thin channels with micro-structured porous walls},
  author={Ling, B. and Oostrom, M. and Tartakovsky, A. M. and Battiato, I.},
  journal={Physics of Fluids},
  volume={30},
  number={7},
  year={2018},
  publisher={AIP Publishing},
  doi={10.1063/1.5031776}
}

@article{bolster2009PoF,
  pages={056601},
  title={Solute dispersion in channels with periodically varying apertures},
  author={Bolster, D. and Dentz, M. and Le Borgne, T},
  journal={Physics of Fluids},
  volume={21},
  number={5},
  year={2009},
  publisher={AIP Publishing},
  doi={10.1063/1.3131982}
}

@article{stroock2000patterning,
  title={Patterning electro-osmotic flow with patterned surface charge},
  author={Stroock, Abraham D. and Weck, Marcus and Chiu, Daniel T. and Huck, Wilhelm T. S. and Kenis, Paul J. A. and Ismagilov, Rustem F. and Whitesides, George M.},
  journal={Physical Review Letters},
  volume={84},
  number={15},
  pages={3314},
  year={2000},
  publisher={APS},
  doi={10.1103/PhysRevLett.84.3314}
}

@article{nishimura1995mass,
  title={Mass transfer enhancement in a symmetric sinusoidal wavy-walled channel for pulsatile flow},
  author={Nishimura, Tatsuo and Kojima, Naoya},
  journal={International Journal of Heat and Mass Transfer},
  volume={38},
  number={9},
  pages={1719--1731},
  year={1995},
  publisher={Elsevier},
  doi={10.1016/0017-9310(94)00275-Z}
}

@article{patera1986exploiting,
  title={Exploiting hydrodynamic instabilities. {R}esonant heat transfer enhancement},
  author={Patera, A. T. and Miki{\'c}, B. B.},
  journal={International Journal of Heat and Mass Transfer},
  volume={29},
  number={8},
  pages={1127--1138},
  year={1986},
  publisher={Elsevier},
  doi={10.1016/0017-9310(86)90144-4}
}

@article{shrestha2025universal,
  title={Universal behaviour in boundary-driven electrokinetic flows},
  author={Shrestha, Ahis and Kirkinis, Eleftherios and Olvera de la Cruz, Monica},
  journal={Journal of Fluid Mechanics},
  volume={1010},
  pages={A50},
  year={2025},
  publisher={Cambridge University Press},
  doi={10.1017/jfm.2025.288}
}

@article{cahill2004electro,
  title={Electro-osmotic streaming on application of traveling-wave electric fields},
  author={Cahill, Brian P and Heyderman, Laura J and Gobrecht, Jens and Stemmer, Andreas},
  journal={Physical Review E},
  volume={70},
  number={3},
  pages={036305},
  year={2004},
  publisher={APS},
  doi={10.1103/PhysRevE.70.036305}
}

@article{ramos1999ac,
  pages={420--422},
  title={{AC} electric-field-induced fluid flow in microelectrodes},
  author={Ramos, Antonio and Morgan, Hywel and Green, Nicolas G and Castellanos, Antonio},
  journal={Journal of Colloid and Interface Science},
  volume={217},
  number={2},
  year={1999},
  doi={10.1006/jcis.1999.6346}
}

@article{shrestha2025self,
  title={Self-generated electrokinetic flows from active-charged boundary patterns},
  author={Shrestha, Ahis and Kirkinis, Eleftherios and Olvera de la Cruz, Monica},
  journal={Physical Review Research},
  volume={7},
  number={2},
  pages={023223},
  year={2025},
  publisher={APS},
  doi={10.1103/PhysRevResearch.7.023223}
}

@article{mortensen2005electrohydrodynamics,
  title={Electrohydrodynamics of binary electrolytes driven by modulated surface potentials},
  author={Mortensen, Niels Asger and Olesen, Laurits H{\o}jgaard and Belmon, Lionel and Bruus, Henrik},
  journal={Physical Review E},
  volume={71},
  number={5},
  pages={056306},
  year={2005},
  publisher={APS},
  doi={10.1103/PhysRevE.71.056306}
}

@article{ajdari1995electro,
  title={Electro-osmosis on inhomogeneously charged surfaces},
  author={Ajdari, Armand},
  journal={Physical Review Letters},
  volume={75},
  number={4},
  pages={755},
  year={1995},
  publisher={APS},
  doi={10.1103/PhysRevLett.75.755}
}

@article{ajdari1996generation,
  title={Generation of transverse fluid currents and forces by an electric field: electro-osmosis on charge-modulated and undulated surfaces},
  author={Ajdari, Armand},
  journal={Physical Review E},
  volume={53},
  number={5},
  pages={4996},
  year={1996},
  publisher={APS},
  doi={10.1103/PhysRevE.53.4996}
}

@article{de2012flow,
  pages={L08404},
  title={Flow topology and scalar mixing in spatially heterogeneous flow fields},
  author={de Barros, F. P. J. and Dentz, M. and Koch, J. and Nowak, W.},
  journal={Geophysical Research Letters},
  volume={39},
  number={8},
  year={2012},
  publisher={Wiley Online Library},
  doi={10.1029/2012GL051302}
}

@article{okubo1970horizontal,
  title={Horizontal dispersion of floatable particles in the vicinity of velocity singularities such as convergences},
  author={Okubo, Akira},
  journal={Deep Sea Research and Oceanographic Abstracts},
  volume={17},
  number={3},
  pages={445--454},
  year={1970},
  publisher={Elsevier},
  doi={10.1016/0011-7471(70)90059-8}
}

@article{marbach2018transport,
  title={Transport and dispersion across wiggling nanopores},
  author={Marbach, Sophie and Dean, David S and Bocquet, Lyd{\'e}ric},
  journal={Nature Physics},
  volume={14},
  number={11},
  pages={1108--1113},
  year={2018},
  publisher={Nature Publishing Group UK London},
  doi={10.1038/s41567-018-0239-0}
}

@article{karnik2005electrostatic,
  doi={10.1021/nl050493b},
  title={Electrostatic control of ions and molecules in nanofluidic transistors},
  author={Karnik, Rohit and Fan, Rong and Yue, Min and Li, Deyu and Yang, Peidong and Majumdar, Arun},
  journal={Nano Letters},
  volume={5},
  number={5},
  pages={943--948},
  year={2005},
  publisher={ACS Publications}
}

@article{daiguji2004ion,
  doi={10.1021/nl0348185},
  title={Ion transport in nanofluidic channels},
  author={Daiguji, Hirofumi and Yang, Peidong and Majumdar, Arun},
  journal={Nano Letters},
  volume={4},
  number={1},
  pages={137--142},
  year={2004},
  publisher={ACS Publications}
}

@article{marbach2019active,
  title={Active control of dispersion within a channel with flow and pulsating walls},
  author={Marbach, Sophie and Alim, Karen},
  journal={Physical Review Fluids},
  volume={4},
  number={11},
  pages={114202},
  year={2019},
  publisher={APS},
  doi={10.1103/PhysRevFluids.4.114202}
}

@article{siwy2002fabrication,
  title={Fabrication of a synthetic nanopore ion pump},
  author={Siwy, Z and Fuli{\'n}ski, A},
  journal={Physical Review Letters},
  volume={89},
  number={19},
  pages={198103},
  year={2002},
  publisher={APS},
  doi={10.1103/PhysRevLett.89.198103}
}

@article{stein2004surface,
  title={Surface-charge-governed ion transport in nanofluidic channels},
  author={Stein, Derek and Kruithof, Maarten and Dekker, Cees},
  journal={Physical Review Letters},
  volume={93},
  number={3},
  pages={035901},
  year={2004},
  publisher={APS},
  doi={10.1103/PhysRevLett.93.035901}
}

@article{petersen2024toward,
  title={Toward Modeling the Structure of Electrolytes at Charged Mineral Interfaces Using Classical Density Functional Theory},
  author={Petersen, Thomas},
  journal={The Journal of Physical Chemistry B},
  volume={128},
  number={16},
  pages={3981--3996},
  year={2024},
  publisher={ACS Publications},
  doi={10.1021/acs.jpcb.3c08045}
}

@article{anderson1985electroosmosis,
  title={Electroosmosis through pores with nonuniformly charged walls},
  author={Anderson, John L and Keith Idol, W},
  journal={Chemical Engineering Communications},
  volume={38},
  number={3-6},
  pages={93--106},
  year={1985},
  publisher={Taylor \& Francis},
  doi={10.1080/00986448508911300}
}

@misc{SM,
  note = {See Supplemental Material at \textcolor{red}{[URL will be inserted by publisher]} for the
          domain transformation, the solution algorithms, and Figs.~S1--S4, which
          includes Refs.~\cite{thompson1982boundary,lele1992compact,ferziger2019computational}}
}

\end{document}